\documentclass[12pt]{article}

\usepackage[totalwidth=460truept,totalheight=600truept]{geometry}
\usepackage{latexsym,graphicx,amsfonts,amssymb,amsmath}
\usepackage[hypertexnames=false,hidelinks]{hyperref}

\global\arraycolsep=1truept

\begin{document}

\null

\vskip1truecm

\begin{center}
{\huge \textbf{Cosmic Inflation as a}}

\vskip.9truecm

{\huge \textbf{Renormalization-Group Flow:}}

\vskip.8truecm

{\huge \textbf{the Running of Power Spectra}}

\vskip.8truecm

{\huge \textbf{in Quantum Gravity}}

\vskip1truecm

\textsl{Damiano Anselmi}

\vskip .1truecm

\textit{Dipartimento di Fisica \textquotedblleft Enrico Fermi", Universit%
\`{a} di Pisa}

\textit{Largo B. Pontecorvo 3, 56127 Pisa, Italy}

\textit{and INFN, Sezione di Pisa,}

\textit{Largo B. Pontecorvo 3, 56127 Pisa, Italy}

damiano.anselmi@unipi.it

\vskip1truecm

\textbf{Abstract}
\end{center}

We study the running of power spectra in inflationary cosmology by
reformulating the slow roll expansion as a renormalization-group flow from
the de Sitter fixed point. The beta function is provided by the equations of
the background metric. The spectra of the scalar and tensor fluctuations
obey RG evolution equations with vanishing anomalous dimensions in the
superhorizon limit. By organizing the perturbative expansion in terms of
leading and subleading logs, we calculate the spectral indices, their
runnings, the runnings of the runnings, etc., to the next-to-leading log
order in quantum gravity with fakeons (i.e., the theory $R+R^{2}+C^{2}$ with
the fakeon prescription/projection for $C^{2}$). We show that these
quantities are related to the spectra in a universal way. We also compute
the first correction to the relation $r=-8n_{T}$ and provide a number of
quantum gravity predictions that can be hopefully tested in the forthcoming
future.

\vfill\eject

\section{Introduction}

\label{intro}\setcounter{equation}{0}

Gravity is the only interaction of nature that permeates the whole universe,
from the smallest distances to the largest ones. Thanks to this, it may
allow us to establish a connection between high-energy physics, specifically
quantum field theory, and cosmology. In this paper we lay out an ingredient
of this relation by reformulating the history of the early universe,
starting from inflation \cite%
{englert,starobinsky,kazanas,sato,guth,linde,steinhardt,linde2} and the
primordial quantum fluctuations \cite{inflation2}, as the evolution of a
peculiar renormalization-group (RG)\ flow in quantum gravity, which we call
\textquotedblleft cosmic RG flow\textquotedblright . The reformulation
offers an alternative way to view the slow roll expansion. Although the
cosmic flow is not originated by the radiative corrections, but by the
dependence on the background\ metric, it resembles the RG flow of quantum
field theory in several respects. For example, the power spectra of the
fluctuations satisfy a Callan-Symanzik equation in the superhorizon limit.

The effects of quantum gravity are expected to become important at energies
that are too large for our laboratories. If we want to test predictions, a
better idea is to use the universe itself as a laboratory. In this context,
a consistent theory of quantum gravity relating the smallest and the largest
scales of magnitude may be of great help. Because the universe was
\textquotedblleft small\textquotedblright\ at the beginning of time, what
happened then may be calculable perturbatively. Since the primordial
evolution of the universe left detectable remnants in the cosmic microwave
background radiation as we see it today, by scrutinizing the sky we may have
chances to put quantum gravity to a test.

Pursuing the idea of a connection between cosmology and high-energy physics,
a sound proposal for quantum gravity should follow from principles similar
to those that lead to the standard model of particle physics, which are
locality, renormalizability and unitarity. The advantage of an approach like
this is that, if it offers an answer, it gives a very constrained, basically
unique one.

A proposal with the claimed features has indeed become available recently
from quantum field theory \cite{LWgrav}. The basic idea is the same that
lead to the introduction of the intermediate W and Z bosons and gave birth
to the standard model of elementary particles.

\begin{figure}[t]
\qquad\qquad\quad\includegraphics[width=10truecm]{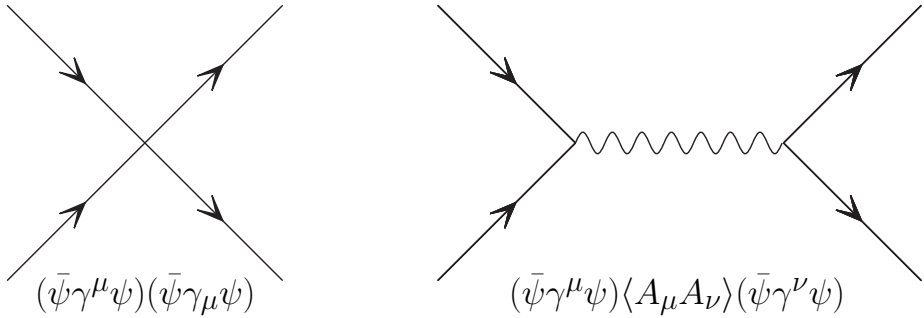}
\caption{Fermi theory of the weak interactions vs standard model}
\label{Fermi}
\end{figure}

At that time, the goal was to remedy the problems of the Fermi theory of the
weak interactions and gain renormalizability while preserving locality and
unitarity. The Fermi theory can be schematically represented by means of
four fermion (current-current) vertices, which are not renormalizable.
Postulating the existence of suitable intermediate bosons, the four fermion
vertices can be effectively generated through the exchange of vector fields,
as shown in figure \ref{Fermi}. In the case of quantum gravity, the left
drawing is replaced by multi graviton vertices and the right drawing is
replaced by trees where the graviton external legs exchange a new type of
particle, the fakeon \cite{fakeons}. 

The difference between the standard model and quantum gravity is that the
intermediate bosons that must be introduced to make sense of the latter are
of a new type, which had to be uncovered anew. The fakeon is a purely
virtual particle, which mediates interactions but does not belong to the
spectrum of asymptotic states. It can be introduced by means of a new
quantization prescription, alternative to the Feynman $i\epsilon $ one,
combined with a projection that is consistent with it.

The prescription amounts to starting from the Euclidean theory and ending
the Wick rotation nonanalytically by means of the average continuation,
anytime a threshold involves fakeons \cite{LWgrav}. The average continuation
is the arithmetic average of the two analytic continuations that circumvent
the threshold \cite{LWFormulation,fakeons}. The computations of
renormalization constants, cross sections, widths and absorptive parts in
quantum gravity with fakeons \cite{UVQG,Absograv} show that the
renormalization is unaffected by the fakeon prescription \cite{fakeons} and
so coincides with the one of the Euclidean version of the theory \cite%
{betaHD}; instead, the absorptive parts are crucially different, since
unitarity depends on them.

The projection works to some extent like the one involved in the
quantization of gauge theories (which is the one that allows us to
consistently drop the gauge-trivial modes and the Faddeev-Popov ghosts from
the spectrum), but does not follow from a symmetry principle. The
combination of the two operations (prescription and projection) is
consistent and returns a theory that is unitary at the fundamental level
(see \cite{fakeons} for the proof to all orders and \cite{LWUnitarity} for
the analysis of the bubble and triangle diagrams).

In cosmology, new aspects play important roles. Because the loop corrections
are negligible up to a high degree of precision \cite{weinberg}, it is
natural to try and quantize the classical limit directly. However, when a
theory contains fake particles, its classical limit is nontrivial. It is not
described by the \textquotedblleft classical\textquotedblright\ action one
starts with to formulate the quantum field theory, because that action is
unprojected \cite{classicization}. Moreover, a purely virtual particle
cannot be found by quantizing something classical \cite{wheelerons}. The way
out is to classicize quantum gravity as explained in \cite{classicization}.
Once the correct classical limit is found, it can be used as the starting
point to study cosmology, as shown by Bianchi, Piva and the current author
in \cite{ABP}.

In this paper we make a step forward towards a better definition of the
relation between cosmology and high-energy physics. We describe the history
of the early universe, starting from the slow roll expansion in inflationary
cosmology, as an evolution that resembles the one of the running couplings
in quantum field theory, although it is essentially different from it,
because it originates from the dependence on the background metric rather
than the radiative corrections. The formal similarities we find about the
two types of flows are quite appealing and might lead to unforseen
developments. We use the resulting setup to derive new quantum gravity
predictions about the spectra and their running behaviors.

Specifically, the cosmic RG flow starts from a fixed point, which is de
Sitter space, and is described by a coupling \textquotedblleft
constant\textquotedblright\ $\alpha $ and its beta function $\beta _{\alpha
}(\alpha )$, which follows from the equations satisfied by the background
metric and can be expressed as a power series in $\alpha $. Around $\alpha
\sim 0$ the beta function is negative and proportional to $\alpha ^{2}$. It
further vanishes for $\alpha =1$, which is however not a fixed point. The
power\ spectra of the scalar and tensor fluctuations obey standard
Callan-Symanzik RG evolution equations in the superhorizon limit, with
vanishing anomalous dimensions.

The tools provided by high-energy physics allow us to gain insight into the
structure of the power spectra, where we can distinguish a core part and an
RG part. The core information is a power series in the coupling $\alpha $
with constant coefficients,\ but has no a priori relation to the beta
function. Because of this, it is specific of the (scalar, tensor) spectrum
we are considering. The RG part of the spectrum, on the other hand, is the
one controlled by the RG equation. For this reason, it is universal (i.e.,
the same for every spectrum) and encoded into the beta function.

The running can be studied efficiently by organizing the perturbative
expansion in terms of leading and subleading logs and resumming all the
powers of $\alpha _{\ast }\ln (k_{\ast }/k)$, where $k$ is the scale, $%
k_{\ast }$ is the pivot scale and $\alpha _{\ast }$ is the value of the
coupling at the pivot scale. In the next sections, we work out the scalar
and tensor spectra and their runnings in quantum gravity to the
next-to-leading log order. We also compute the first correction to the
relation $r+8n_{T}=0$, where $r$ is the tensor-to-scalar ratio and $n_{T}$
is the tensor tilt. These are quantities that can be hopefully measured in
the future.

As already stressed, the cosmic RG flow is originated by the dependence on
the background metric. In this sense, it is the intrinsic flow of inflation.
The expression \textquotedblleft running\textquotedblright ,\ widely used
for the dependence of the tilts on $\ln k$ \cite{Planck18}, fits the
terminology used here. Instead, the factor $z$ that relates the curvature
perturbation $\mathcal{R}$ to the Mukhanov-Sasaki variable (see for example 
\cite{baumann}) is not interpreted as a wave-function renormalization
constant $Z^{1/2}$. If it were so, it would lead to an anomalous dimension,
but we find that the anomalous dimensions vanish. Other concepts that can be
found in the literature have meanings that differ from the ones we attribute
to them. For example, the beta function $\beta _{\alpha }$ defined in this
paper does not match the quantity called \textquotedblleft beta
function\textquotedblright\ by Bin\'{e}truy, Kiritsis, Mabillard, Pieroni
and Rosset\ in \cite{altrabeta} (which is proportional to the quantity $%
\alpha $ that we call \textquotedblleft coupling\textquotedblright ). We
think that our interpretation of the formal relation between cosmology and
high-energy physics is more to the point than the one proposed there,
because it allows us to dissect the structure of the power spectra.

Although the leading log expansion we are talking about has nothing to do
with the leading logs due to radiative corrections in quantum field theory,
it is interesting to incorporate those effects, such as the running of
masses and coupling constants induced by high-energy physics, in the
investigation of primordial cosmology. This was done for example in ref. 
\cite{odintsov} by Myrzakulov, Odintsov and Sebastiani. It is also worth to
mention the RG induced inflation defined by M\'{a}ri\'{a}n, Defenu,
Jentschura, Trombettoni and N\'{a}ndori in ref. \cite{induced}. For various
applications it might be useful to combine the different types of flows
together.

In models without fakeons, the running of spectral indices has been
calculated in various scenarios \cite{run1}, including subleading
corrections \cite{run2}. Besides upgrading the techniques of \cite{run1,run2}
and including purely virtual quanta, the understanding offered in this paper
allows us to appreciate important aspects of inflation and make calculations
more efficiently.

Hopefully, primordial cosmology will provide an arena for precision tests of
quantum gravity. With this in mind, we use the results of our calculations
to derive a number of predictions that might be tested experimentally in the
forthcoming years.

The paper is organized as follows. In section \ref{betaf} we study inflation
as an RG flow, define the coupling $\alpha $ and study its beta function. We
also organize the perturbative expansion by resumming the leading logs and
the subleading logs, and use these tools to compute the running coupling to
various orders. In section \ref{R2} we study the scalar and tensor spectra
in the limit of infinitely heavy fakeon, which returns the Starobinsky $%
R+R^{2}$ theory \cite{starobinsky,vilenkin}. In section \ref{qg} we upgrade
the results of section \ref{R2} to quantum gravity, emphasizing the
dependence on the fakeon mass $m_{\chi }$. In section \ref{predictions} we
summarize the predictions, while section \ref{conclusions} contains the
conclusions.

\section{The cosmic RG\ flow}

\label{betaf}\setcounter{equation}{0}

In this section we define the cosmic RG flow and study the running coupling $%
\alpha $ and its beta function $\beta _{\alpha }$. For later use, we
organize and resum the perturbative expansion in terms of leading logs and
subleading logs.

Quantum gravity with fakeons is described by a triplet made of the graviton,
a massive scalar $\phi $ (the inflaton) and a massive spin-2 fakeon $\chi
_{\mu \nu }$. The starting, unprojected classical action can be written in
the form%
\begin{equation}
S_{\text{QG}}=-\frac{1}{16\pi G}\int \mathrm{d}^{4}x\sqrt{-g}\left( R+\frac{1%
}{2m_{\chi }^{2}}C_{\mu \nu \rho \sigma }C^{\mu \nu \rho \sigma }\right) +%
\frac{1}{2}\int \mathrm{d}^{4}x\sqrt{-g}\left( D_{\mu }\phi D^{\mu }\phi
-2V(\phi )\right) ,  \label{sqgeq}
\end{equation}%
where $V(\phi )$ is the Starobinsky potential%
\begin{equation}
V(\phi )=\frac{m_{\phi }^{2}}{2\hat{\kappa}^{2}}\left( 1-\mathrm{e}^{\hat{%
\kappa}\phi }\right) ^{2},  \label{staropote}
\end{equation}%
while $\hat{\kappa}=\sqrt{16\pi G/3}$ and $m_{\phi }$, $m_{\chi }$ are the
masses of $\phi $ and $\chi _{\mu \nu }$, respectively. For convenience, we
have switched off both the cosmological term and the matter sector.

The theory (\ref{sqgeq}) is renormalizable. Indeed, up to a standard,
nonderivative field redefinition, it is equivalent to the action%
\begin{equation}
\tilde{S}_{\text{QG}}(g,\Phi )=-\frac{1}{16\pi G}\int \mathrm{d}^{4}x\sqrt{-g%
}\left( R+\frac{1}{2m_{\chi }^{2}}C_{\mu \nu \rho \sigma }C^{\mu \nu \rho
\sigma }-\frac{R^{2}}{6m_{\phi }^{2}}\right) ,  \label{Sgeom}
\end{equation}%
which is renormalizable by power counting \cite{stelle}, once the
cosmological term is reinstated. Both (\ref{sqgeq}) and (\ref{Sgeom})
contain the square $C^{2}$ of the Weyl tensor $C_{\mu \nu \rho \sigma }$. If
all the poles of the free propagators (in the expansion of the metric around
flat space) are quantized by means of the Feynman $i\epsilon $ prescription,
Stelle's theory \cite{stelle} is obtained, which contains a spin-2 ghost and
violates unitarity.

The solution is to quantize the massive spin-2 pole by means of the fakeon
prescription, which allows us to project the particle away from the physical
spectrum at the fundamental level. The procedure works in cosmology under
the consistency condition $m_{\chi }>m_{\phi }/4$, which puts a lower bound
on the mass of the fakeon with respect to the mass of the inflaton \cite{ABP}%
.

Throughout this paper, we work in the \textquotedblleft inflaton
framework\textquotedblright\ and conform to the notation of \cite{ABP}. The
inflaton framework is the one where the scalar field $\phi $ is introduced
explicitly and the action is (\ref{sqgeq}). The action (\ref{Sgeom}) defines
the \textquotedblleft geometric framework\textquotedblright\ of ref. \cite%
{ABP}. It can be shown that the two frameworks lead to the same physical
results, as expected, although several intermediate steps may look rather
different. It is worth to stress that, even if we use (\ref{sqgeq}), we do
not have the freedom to change the potential $V$ of (\ref{staropote}),
because if we did so we would destroy the renormalizability of the theory.

To define the cosmic RG flow, we start from the Friedmann equations, which
are unaffected by the $C^{2}$ term and read%
\begin{equation}
\dot{H}=-4\pi G\dot{\phi}^{2},\qquad H^{2}=\frac{4\pi G}{3}\left( \dot{\phi}%
^{2}+2V(\phi )\right) ,\qquad \ddot{\phi}+3H\dot{\phi}=-V^{\prime }(\phi ),
\label{frie}
\end{equation}%
where $H=\dot{a}/a$ is the Hubble parameter. We study the evolution from the
de Sitter limit (which is $\phi \rightarrow -\infty $). For the applications
we have in mind it is sufficient to cover the region $\phi <0$, so the
equations we write below are valid in that domain, which means to the left
of the minimum $\phi =0$ of the potential (\ref{staropote}). They can be
extended to the other domains by flipping the signs in front of the square
roots we are going to meet.

It is useful to define the \textquotedblleft coupling\textquotedblright 
\begin{equation}
\alpha =\sqrt{\frac{4\pi G}{3}}\frac{\dot{\phi}}{H}=\sqrt{-\frac{\dot{H}}{%
3H^{2}}}  \label{alf}
\end{equation}%
and eliminate $V$ and $\dot{\phi}$ by means of the first two equations of (%
\ref{frie}). So doing, we obtain%
\begin{equation}
\dot{\phi}=\sqrt{\frac{3}{4\pi G}}\alpha H,\qquad V=\frac{3}{8\pi G}%
(1-\alpha ^{2})H^{2}.  \label{faH}
\end{equation}%
Note that we must have $-1\leqslant \alpha \leqslant 1$. Moreover, in the
domain we are interested in we can take $\alpha $ to be nonnegative. Then
the flow describes the evolution from $\phi =-\infty $ to $\phi =0$.

We eliminate $\ddot{\phi}$ from the last equation of (\ref{frie}), by noting
that the potential (\ref{staropote}) satisfies%
\begin{equation}
V^{\prime }=4V\sqrt{\frac{4\pi G}{3}}-m_{\phi }\sqrt{2V}.  \label{vip}
\end{equation}%
Differentiating the first equation of (\ref{faH}) with respect to time and
using the last equation (\ref{frie}), followed by (\ref{alf}), (\ref{vip})
and the second of (\ref{faH}), we find 
\begin{equation}
\dot{\alpha}=m_{\phi }\sqrt{1-\alpha ^{2}}-H(2+3\alpha )\left( 1-\alpha
^{2}\right) .  \label{equa}
\end{equation}%
This equation can be extended to the right of the $V$ minimum ($\phi >0$) by
flipping the sign in front of the square root that appears on its right-hand
side. It can be extended to $\alpha <0$ (i.e., $\dot{\phi}<0$), by flipping
the sign in front of the square root of (\ref{alf}).

Equation (\ref{equa}) still contains $H$, which can eliminated as follows.
We first write the most general expansion 
\begin{equation}
H=\sum_{n=0}^{\infty }h_{n}\alpha ^{n}  \label{posit}
\end{equation}%
in powers of $\alpha $, where $h_{n}$ are unknown constants. Then we
differentiate this relation with respect to time. The left-hand side gives $%
\dot{H}=-3\alpha ^{2}H^{2}$, by (\ref{alf}), which is turned into a power
series in $\alpha $ by using (\ref{posit}) again. The right-hand side is
turned into another power series in $\alpha $ by means of (\ref{equa}) and (%
\ref{posit}). Comparing the two sides, we find 
\begin{equation}
H=\frac{m_{\phi }}{2}\left[ 1-\frac{3\alpha }{2}+\frac{7\alpha ^{2}}{4}-%
\frac{47\alpha ^{3}}{24}+\frac{293\alpha ^{4}}{144}+\mathcal{O}(\alpha ^{5})%
\right] .  \label{acca}
\end{equation}%
The formula can be pushed to arbitrarily high orders.

Thanks to (\ref{acca}), we can use (\ref{equa}) to express $\dot{\alpha}$ as
a power series in $\alpha $. However, the derivative is still with respect
to time. It is convenient to introduce the conformal time%
\begin{equation}
\tau =-\int_{t}^{+\infty }\frac{\mathrm{d}t^{\prime }}{a(t^{\prime })}
\label{tau}
\end{equation}%
and convert the derivative by means of 
\begin{equation*}
\frac{\mathrm{d}}{\mathrm{d}t}=\frac{H}{aH\tau }\frac{\mathrm{d}}{\mathrm{%
d\ln }|\tau |}.
\end{equation*}%
For this purpose, we need the expansion%
\begin{equation}
-aH\tau =1+3\alpha ^{2}+12\alpha ^{3}+91\alpha ^{4}+\mathcal{O}(\alpha ^{5}),
\label{aHtau}
\end{equation}%
which can be derived with the method used for (\ref{acca}).

At this point, it is straightforward to convert (\ref{equa}) into the beta
function%
\begin{equation}
\beta _{\alpha }\equiv \frac{\mathrm{d}\alpha }{\mathrm{d\ln }|\tau |}%
=-2\alpha ^{2}f(\alpha ),  \label{beta}
\end{equation}%
where%
\begin{equation}
f(\alpha )=1+\frac{5}{6}\alpha +\frac{25}{9}\alpha ^{2}+\frac{383}{27}\alpha
^{3}+\frac{8155}{81}\alpha ^{4}+\mathcal{O}(\alpha ^{5}).  \label{falfa}
\end{equation}%
Note that $\beta _{\alpha }$ is negative for $\alpha $ small, so the cosmic
RG flow is \textquotedblleft asymptotically free\textquotedblright , which
means that it gives the de Sitter metric in the infinite-past limit $%
t\rightarrow -\infty $, which is also $\tau \rightarrow -\infty $.

It is easy to show that $\alpha =0$ is the only fixed point of the flow.
Indeed, a fixed point has $\alpha =$ constant. The right-hand side of
equation (\ref{equa}) vanishes for 
\begin{equation}
\frac{m_{\phi }}{2}=\pm H\left( 1+\frac{3}{2}\alpha \right) \sqrt{1-\alpha
^{2}}  \label{fix}
\end{equation}%
(the $\pm $ being inserted to cover the case $\phi >0$) and for $\alpha =\pm
1$. We consider the two cases separately.

Equation (\ref{fix}) implies $\dot{H}=0$. However, $\dot{H}$ is also equal
to $-3\alpha ^{2}H^{2}$, so we can either have $\alpha =0$ or $H=0$. The
option $H=0$ is not acceptable, because it is incompatible with (\ref{fix}).
In the end, we obtain the de Sitter fixed point $\alpha =0$, with $H=m_{\phi
}/2$ (since $H$ cannot be negative).

As far as the cases $\alpha =\pm 1$ are concerned, they are zeros of the
beta function, but they are not true fixed points. Indeed, for $|\alpha
|\lesssim 1$, (\ref{equa}) gives 
\begin{equation*}
\dot{\alpha}\sim \pm m_{\phi }\sqrt{1-\alpha ^{2}}.
\end{equation*}%
This equation is solved by $\alpha \sim \cos (m_{\phi }(t-t_{0}))$, which
takes us to the oscillating behavior of the reheating phase. Moreover, the
spurious zeros at $\alpha =\pm 1$ disappear with a simple reparametrization,
such as $\alpha =\sin \theta $.

To simplify some calculations, it is convenient to define the variable $\eta
=-k\tau $, where $k$ is just an arbitrary constant for the moment. We solve (%
\ref{beta}) by writing 
\begin{equation}
\mathrm{d\ln }\eta =-\frac{\mathrm{d}\alpha }{2\alpha ^{2}}\left( 1-\frac{5}{%
6}\alpha -\frac{25}{12}\alpha ^{2}-\frac{2189}{216}\alpha ^{3}+\mathcal{O}%
(\alpha ^{4})\right)  \label{dleta}
\end{equation}%
and integrating term by term. The solution can be organized by means of
leading and subleading logs, which can be easily resummed.

Let $\alpha _{0}$ denote the value of $\alpha $ at $\eta =1$. The expansion
in terms of leading and subleading logs is the expansion in powers of $%
\alpha _{0}$ under the assumption that $\alpha _{0}\ln \eta $ is of order
unity. This means that the powers of $\alpha _{0}\ln \eta $ must be resummed
into exact expressions. Modulo overall factors $\alpha _{0}$, the leading
logs are the powers $\alpha _{0}^{n}\ln ^{n}\eta $, $n\geqslant 0$, the
next-to-leading logs are the corrections proportional to $\alpha
_{0}^{n+1}\ln ^{n}\eta $, $n\geqslant 0$, the next-to-next-to-leading logs
are the corrections $\alpha _{0}^{n+2}\ln ^{n}\eta $, and so on.

Thanks to the renormalization group, the inclusion of each set of
corrections requires to add just one term of the expansion in powers of $%
\alpha $. At the leading log level, it is sufficient to keep the first term
inside the parenthesis of (\ref{dleta}), which gives the running coupling%
\begin{equation}
\alpha (-\tau )=\frac{\alpha _{0}}{1+2\alpha _{0}\ln \eta }.  \label{lead}
\end{equation}%
The expansion in powers of $\alpha _{0}\ln \eta $ is just a geometric series
in this case. We can view the initial condition $\alpha _{0}$ as implicitly $%
k$ dependent, i.e., as the coupling%
\begin{equation}
\alpha _{0}=\alpha (1/k)  \label{runna}
\end{equation}%
at $|\tau |=1/k$. Then $\alpha (-\tau )$ is $k$ independent, i.e., it is the
coupling at conformal time $\tau $.

More generally, the running coupling can be organized in the form 
\begin{equation}
\alpha (-\tau )=\frac{\alpha _{0}}{\lambda }\prod_{n=1}^{\infty }(1+\alpha
_{0}^{n}\gamma _{n}(\lambda )),\qquad \lambda \equiv 1+2\alpha _{0}\ln \eta .
\label{alfa}
\end{equation}%
The first few functions $\gamma _{n}(\lambda )$ are%
\begin{eqnarray}
\gamma _{1}(\lambda ) &=&-\frac{5\ln \lambda }{6\lambda },\qquad \gamma
_{2}(\lambda )=\frac{25}{12\lambda ^{2}}\left[ 1-\lambda -\frac{\ln \lambda 
}{3}(1-\ln \lambda )\right] ,  \notag \\
\gamma _{3}(\lambda ) &=&\frac{1}{432\lambda ^{3}}\left[ (1-\lambda
)(2939+2189\lambda )-125(14-6\lambda -3\ln \lambda )\ln \lambda \right] .
\label{gamma}
\end{eqnarray}

To give an example, the running coupling to the next-to-leading log order
can be found by keeping the first two terms inside the parenthesis of (\ref%
{dleta}), which gives%
\begin{equation}
\alpha (-\tau )=\frac{\alpha _{0}}{1+2\alpha _{0}\ln \eta }\left( 1-\frac{%
5\alpha _{0}}{6}\frac{\ln (1+2\alpha _{0}\ln \eta )}{1+2\alpha _{0}\ln \eta }%
\right) .  \label{runca}
\end{equation}

\section[Limit of heavy fakeon]{Limit of heavy fakeon ($m_{\chi }=\infty $,
action $R+R^{2}$)}

\label{R2}\setcounter{equation}{0}

In this section, we study the cosmic RG flow in the limit of infinitely
heavy fakeon, $m_{\chi }\rightarrow \infty $. The action (\ref{sqgeq}) turns
into%
\begin{equation}
S_{\text{QG}}=-\frac{1}{16\pi G}\int \mathrm{d}^{4}x\sqrt{-g}R+\frac{1}{2}%
\int \mathrm{d}^{4}x\sqrt{-g}\left( D_{\mu }\phi D^{\mu }\phi -2V(\phi
)\right) .  \label{staroac}
\end{equation}%
Recalling that the potential is (\ref{staropote}), we obtain the Starobinsky 
$R+R^{2}$ theory in the inflaton approach, which provides a good arena to
illustrate the strategy of our calculations before generalizing them to
quantum gravity.

We show that the spectra of the tensor and scalar fluctuations obey RG
evolution equations in the superhorizon limit, with vanishing anomalous
dimensions. Due to this, the tilts and the running coefficients are related
to the spectra in a universal way. The procedure we describe can be extended
to arbitrarily high orders. We compute the first few orders explicitly.

For reviews that contain details on the parametrizations of the metric
fluctuations and their transformations under diffeomorphisms, see \cite%
{baumann,reviews}.

\subsection{Tensor fluctuations}

\label{tensor}

To study the tensor fluctuations, we parametrize the metric as 
\begin{equation}
g_{\mu \nu }=\text{diag}(1,-a^{2},-a^{2},-a^{2})-2a^{2}\left( u\delta _{\mu
}^{1}\delta _{\nu }^{1}-u\delta _{\mu }^{2}\delta _{\nu }^{2}+v\delta _{\mu
}^{1}\delta _{\nu }^{2}+v\delta _{\mu }^{2}\delta _{\nu }^{1}\right) ,
\label{met}
\end{equation}%
where $u=u(t,z)$ and $v=v(t,z)$ denote the graviton modes. Denoting the
Fourier transform of $u(t,z)$ with respect to the coordinate $z$ by $u_{%
\mathbf{k}}(t)$, where $\mathbf{k}$ is the space momentum, the quadratic
Lagrangian obtained from (\ref{staroac}) is 
\begin{equation}
(8\pi G)\frac{\mathcal{L}_{\text{t}}}{a^{3}}=\dot{u}_{\mathbf{k}}\dot{u}_{-%
\mathbf{k}}-\frac{k^{2}}{a^{2}}u_{\mathbf{k}}u_{-\mathbf{k}},  \label{lut}
\end{equation}%
plus an identical contribution for $v_{\mathbf{k}}$, where $k=|\mathbf{k}|$.
Defining%
\begin{equation}
w_{\mathbf{k}}=au_{\mathbf{k}}\sqrt{\frac{k}{4\pi G}},  \label{w}
\end{equation}%
and switching to the variable $\eta =-k\tau $, the action reads%
\begin{equation}
S_{\text{t}}=\frac{1}{2}\int \mathrm{d}\eta \left[ w^{\prime 2}-w^{2}+\left(
\nu _{\text{t}}^{2}-\frac{1}{4}\right) \frac{w^{2}}{\eta ^{2}}\right] ,
\label{sred}
\end{equation}%
where the prime denotes the derivative with respect to $\eta $ and%
\begin{equation}
\nu _{\text{t}}^{2}-\frac{1}{4}=2\tau ^{2}a^{2}H^{2}\left( 1-\frac{3}{2}%
\alpha ^{2}\right) .  \label{ntexa}
\end{equation}%
Using (\ref{aHtau}), we find%
\begin{equation}
\nu _{\text{t}}=\frac{3}{2}+3\alpha ^{2}+16\alpha ^{3}+\frac{355}{3}\alpha
^{4}+\mathcal{O}(\alpha ^{5}).  \label{nt}
\end{equation}

For a while, we drop the subscript $\mathbf{k}$, since no confusion is
expected to arise. The equation derived from (\ref{sred}) is%
\begin{equation}
w^{\prime \prime }+w-\left( \nu _{\text{t}}^{2}-\frac{1}{4}\right) \frac{w}{%
\eta ^{2}}=0  \label{weq}
\end{equation}%
and must be solved with the Bunch-Davies vacuum condition%
\begin{equation}
w(\eta )\sim \frac{\mathrm{e}^{i\eta }}{\sqrt{2}}\qquad \text{for large }%
\eta .  \label{bunch}
\end{equation}

The solution can be worked out by expanding in powers of $\alpha _{0}=\alpha
(1/k)$. Since there is no $\mathcal{O}(\alpha )$ term in (\ref{nt}), the
first correction to $w$ is of order $\alpha _{0}^{2}$. We have 
\begin{equation}
w(\eta )=w_{0}(\eta )+\alpha _{0}^{2}w_{2}(\eta )+\alpha _{0}^{3}w_{3}(\eta
)+\cdots .  \label{weta}
\end{equation}%
Inserting (\ref{weta}) into (\ref{weq}), we obtain equations of the form%
\begin{equation}
w_{n}^{\prime \prime }+w_{n}-\frac{2w_{n}}{\eta ^{2}}=\frac{g_{n}(\eta )}{%
\eta ^{2}}.  \label{gn}
\end{equation}%
where $g_{0}(\eta )=g_{1}(\eta )=0$, while $g_{n}(\eta )$ , $n>1$, are
functions that are determined recursively from $w_{j}$ with $j<n$. Using the
expansion (\ref{nt}), then expressing $\alpha $, which is $\alpha (-\tau )$,
as the running coupling of the previous section, which we can expand in
powers of $\alpha _{0}$ by means of formula (\ref{alfa}), we find%
\begin{equation}
g_{2}=9w_{0},\qquad g_{3}=12w_{0}(4-3\ln \eta ),\qquad
g_{4}=9w_{2}+2w_{0}(182-159\ln \eta +54\ln ^{2}\eta ),\quad  \label{g2}
\end{equation}%
etc.

The solution for $n=0$ is%
\begin{equation}
w_{0}=\frac{(\eta +i)}{\eta \sqrt{2}}\mathrm{e}^{i\eta },  \label{w0}
\end{equation}%
where the arbitrary constants are determined by the Bunch-Davies condition (%
\ref{bunch}). The solutions for $n>0$ are 
\begin{equation}
w_{n}(\eta )=\int_{\eta }^{\infty }\frac{g_{n}(\eta ^{\prime })\mathrm{d}%
\eta ^{\prime }}{\eta \eta ^{\prime 3}}\left[ (\eta -\eta ^{\prime })\cos
(\eta -\eta ^{\prime })-(1+\eta \eta ^{\prime })\sin (\eta -\eta ^{\prime })%
\right] .  \label{wn}
\end{equation}%
Again, the arbitrary constants are determined to preserve (\ref{bunch}),
which requires $w_{n}(\eta )\rightarrow 0$ for $\eta \rightarrow \infty $
for every $n>1$.

For example, using (\ref{g2}) for $g_{2}$ and $g_{3}$, we find%
\begin{eqnarray}
w_{2}(\eta ) &=&\frac{3}{\eta \sqrt{2}}\left[ 2i\mathrm{e}^{i\eta }+(\eta -i)%
\mathrm{e}^{-i\eta }\left( \hspace{0.01in}\text{Ei}(2i\eta )-i\pi \right) %
\right] ,  \notag \\
w_{3}(\eta ) &=&-\frac{12i\sqrt{2}\mathrm{e}^{i\eta }\ln \eta }{\eta }+\frac{%
3\sqrt{2}(\eta -i)\mathrm{e}^{-i\eta }}{\eta }\left( 2i\pi -\frac{\pi ^{2}}{%
12}-i\pi \gamma _{M}-2(\ln \eta +1)\text{Ei}(2i\eta )\right.  \notag \\
&&\qquad \qquad +i\pi \ln \eta +(\ln \eta +\gamma _{M})^{2}+4i\eta
\,F_{2,2,2}^{1,1,1}\left( 2i\eta \right) 
\Big)%
,  \label{w23}
\end{eqnarray}%
where Ei is the exponential-integral function, $F_{b_{1},\cdots
,b_{q}}^{a_{1},\cdots ,a_{p}}(z)$ is the generalized hypergeometric function 
$_{p}F_{q}(\{a_{1},\cdots ,a_{p}\},\{b_{1},\cdots ,b_{q}\};z)$ and $\gamma
_{M}\equiv \gamma _{E}+\ln 2$, $\gamma _{E}$ being the Euler-Mascheroni
constant. The combination $\gamma _{M}$ is going to appear frequently from
now on.

In the limit $\eta \rightarrow \infty $, the functions $g_{n}$ and $w_{n}$, $%
n>1$, satisfy%
\begin{equation}
\lim_{\eta \rightarrow \infty }\eta ^{1-\delta }g_{n}(\eta )=0,\qquad
\lim_{\eta \rightarrow \infty }\eta ^{1-\delta }w_{n}(\eta )=0,  \label{inf}
\end{equation}%
for every $\delta >0$.

The power spectrum must be calculated in the superhorizon limit, which is $%
\eta \rightarrow 0$. There, we have%
\begin{equation}
g_{n}(\eta )\underset{\eta \sim 0}{\sim }\frac{1}{\eta }P_{n-2}^{(g)}(\ln
\eta )+\mathcal{O}(\ln ^{n-2}\eta )\qquad w_{n}(\eta )\underset{\eta \sim 0}{%
\sim }\frac{1}{\eta }P_{n-1}^{(w)}(\ln \eta )+\mathcal{O}(\ln ^{n-1}\eta ),
\label{superhor}
\end{equation}%
where $P_{k}^{(g)}$ and $P_{k}^{(w)}$ are polynomials of degree $k$.
Moreover, the superhorizon limit allows us to drop the term $w_{n}$ in
equation (\ref{gn}), because it is dominated by $2w_{n}/\eta ^{2}$. Once we
do that, the solution (\ref{wn}) simplifies to%
\begin{equation}
w_{n}(\eta )\underset{\eta \sim 0}{\sim }-\frac{1}{3\eta }\int_{\eta
_{1}}^{\eta }g_{n}(\eta ^{\prime })\mathrm{d}\eta ^{\prime }+\frac{\eta ^{2}%
}{3}\int_{\eta _{2}}^{\eta }\frac{g_{n}(\eta ^{\prime })\mathrm{d}\eta
^{\prime }}{\eta ^{\prime 3}},  \label{appros}
\end{equation}%
for $n>1$, where $\eta _{1}$ and $\eta _{2}$ are arbitrary constants. In
particular, if we take 
\begin{equation}
g_{n}(\eta )\underset{\eta \sim 0}{\sim }\frac{c_{n}}{\eta }\ln ^{n-2}\eta +%
\text{ subleading,}  \label{cn}
\end{equation}%
where $c_{n}$ are constants, we obtain 
\begin{equation}
w_{n}(\eta )\underset{\eta \sim 0}{\sim }-\frac{c_{n}}{3(n-1)}\frac{\ln
^{n-1}\eta }{\eta }+\text{ subleading,\qquad }n>1\text{.}  \label{wcn}
\end{equation}

Now we use these results to resum the leading log corrections in the power
spectrum. For this purpose it is enough to truncate (\ref{nt}) to order $%
\alpha ^{2}$ and replace $\alpha $ with (\ref{lead}), so equation (\ref{weq}%
) simplifies to 
\begin{equation*}
w^{\prime \prime }+w-\frac{2w}{\eta ^{2}}=\frac{9\alpha _{0}^{2}w_{0}}{\eta
^{2}(1+2\alpha _{0}\ln \eta )^{2}}.
\end{equation*}%
Then we use the expansion (\ref{weta}) to read $g_{n}$ and so the constants $%
c_{n}$ of (\ref{cn}), which turn out to be%
\begin{equation*}
c_{n}=9i(-1)^{n}(n-1)2^{n-3}\sqrt{2}.
\end{equation*}%
Using (\ref{wcn}) and resumming (\ref{weta}), we find the first correction
to $w(\eta )$:%
\begin{equation}
w(\eta )\sim \frac{i}{\eta \sqrt{2}}\left( 1-\frac{3\alpha _{0}^{2}\ln \eta 
}{1+2\alpha _{0}\ln \eta }\right) .  \label{w0tensor}
\end{equation}

We quantize (\ref{sred}) as usual. Reinstating the subscript $\mathbf{k}$,
the operator associated with the fluctuation $u_{\mathbf{k}}$ is%
\begin{equation*}
\hat{u}_{\mathbf{k}}(\tau )=u_{\mathbf{k}}(\tau )\hat{a}_{\mathbf{k}}+u_{-%
\mathbf{k}}^{\ast }(\tau )\hat{a}_{-\mathbf{k}}^{\dagger },
\end{equation*}%
where $\hat{a}_{\mathbf{k}}^{\dagger }$ and $\hat{a}_{\mathbf{k}}$ are
creation and annihilation operators, satisfying $[\hat{a}_{\mathbf{k}},\hat{a%
}_{\mathbf{k}^{\prime }}^{\dagger }]=(2\pi )^{3}\delta ^{(3)}(\mathbf{k}-%
\mathbf{k}^{\prime })$. The power spectrum $\mathcal{P}_{u}$ is defined by%
\begin{equation}
\langle \hat{u}_{\mathbf{k}}(\tau )\hat{u}_{\mathbf{k}^{\prime }}(\tau
)\rangle =(2\pi )^{3}\delta ^{(3)}(\mathbf{k}+\mathbf{k}^{\prime })\frac{%
2\pi ^{2}}{k^{3}}\mathcal{P}_{u},\qquad \mathcal{P}_{u}=\frac{k^{3}}{2\pi
^{2}}|u_{\mathbf{k}}|^{2}.  \label{pu}
\end{equation}%
Summing over the tensor polarizations $u$ and $v$ and converting to the
common normalization, the power spectrum of the tensor fluctuations is%
\begin{equation}
\mathcal{P}_{T}(k)=16\mathcal{P}_{u}(k).  \label{pt0}
\end{equation}%
Using (\ref{w}) and (\ref{w0tensor}), we find%
\begin{equation*}
\mathcal{P}_{u}=\frac{2Gk^{2}}{\pi a^{2}}|w_{\mathbf{k}}|^{2}=\frac{GH^{2}}{%
\pi a^{2}H^{2}\tau ^{2}}\left( 1-\frac{3\alpha _{0}^{2}\ln \eta }{1+2\alpha
_{0}\ln \eta }\right) ^{2}.
\end{equation*}%
Formula (\ref{aHtau}) can be approximated to $-aH\tau =1$ to the order we
are interested in, while\ (\ref{acca}) can be truncated to $H=m_{\phi
}(1-3(\alpha /2))/2$. In the end, we obtain%
\begin{equation}
\mathcal{P}_{T}(k)=\frac{4Gm_{\phi }^{2}}{\pi }\left( 1-3\alpha (-\tau
)\right) \left( 1-\frac{6\alpha _{0}^{2}\ln \eta }{1+2\alpha _{0}\ln \eta }%
\right) =\frac{4Gm_{\phi }^{2}}{\pi }\left( 1-3\alpha (1/k)\right) ,
\label{pt}
\end{equation}%
having used (\ref{lead})\ again and dropped higher orders.

Note that the $\eta $ dependence has disappeared. This is a general fact,
due to the RG evolution equation obeyed by the power spectrum (see below).
The running coupling $\alpha (1/k)$ at $1/k$ can be written as the coupling $%
\alpha _{\ast }$ at a reference scale $1/k_{\ast }$ evolved to $1/k$ by
means of the RG equation.

The spectral index is defined by%
\begin{equation}
n_{T}(k)=\frac{\mathrm{d}\ln \mathcal{P}_{T}(k)}{\mathrm{d}\ln k},
\label{indextenso}
\end{equation}%
hence we have%
\begin{equation}
\mathcal{P}_{T}(k)=\mathcal{P}_{T}(k_{\ast })\exp \left( \int_{k_{\ast }}^{k}%
\frac{\mathrm{d}k^{\prime }}{k^{\prime }}n_{T}(k^{\prime })\right) .
\label{ptens}
\end{equation}%
Viewing $\mathcal{P}_{T}(k)$ and $n_{T}(k)$ as functions $\mathcal{\tilde{P}}%
_{T}(\alpha )$ and $\tilde{n}_{T}(\alpha )\hspace{0.01in}$ of the coupling $%
\alpha $, we can also write%
\begin{equation*}
\mathcal{\tilde{P}}_{T}(\alpha )=\mathcal{\tilde{P}}_{T}(\alpha _{\ast
})\exp \left( -\int_{\alpha _{\ast }}^{\alpha }\frac{\tilde{n}_{T}(\alpha
^{\prime })\hspace{0.01in}\mathrm{d}\alpha ^{\prime }}{\beta _{\alpha
}(\alpha ^{\prime })}\right) .
\end{equation*}

Using (\ref{pt}) and the beta function (\ref{beta}), and dropping
higher-order corrections, we find the first contributions to $n_{T}$ and its
running coefficients:%
\begin{equation*}
n_{T}=\frac{\mathrm{d}\ln \mathcal{P}_{T}}{\mathrm{d}\ln k}=3\beta _{\alpha
}=-6\alpha ^{2},\quad \frac{\mathrm{d}n_{T}}{\mathrm{d}\ln k}=12\alpha \beta
_{\alpha }=-24\alpha ^{3},\quad \frac{\mathrm{d}^{n}n_{T}}{\mathrm{d}\ln k%
\hspace{0.01in}^{n}}=-6\cdot 2^{n}(n+1)!\alpha ^{n+2},
\end{equation*}%
where $\alpha $ now stands for $\alpha (1/k)$. All the coefficients are
related to $\mathcal{P}_{T}$ by the RG equations, which we derive now.

\subsection{RG equation of the power spectrum}

Now we derive the RG evolution equation obeyed by the power spectrum in the
superhorizon limit. We have seen that the running coupling $\alpha (-\tau )$
is an expansion in powers $\alpha _{0}^{l+1}(\alpha _{0}\ln \left\vert \tau
\right\vert )^{n}$ with $l\geqslant 0$, $n\geqslant 0$. Using (\ref{acca})
we find that $H$ is $m_{\phi }/2$ plus an expansion of the same type. By (%
\ref{aHtau}), $-aH\tau $ is equal to 1 plus an analogous expansion. Finally,
integrating (\ref{tau}) we find the expansions 
\begin{eqnarray}
-\frac{m_{\phi }t}{2} &=&\left( 1+\alpha _{0}^{l+1}(\alpha _{0}m_{\phi
}t)^{n}\right) \ln \frac{m_{\phi }|\tau |}{2}=\left( 1+\alpha
_{0}^{l+1}\left( \alpha _{0}\ln \frac{m_{\phi }|\tau |}{2}\right)
^{n}\right) \ln \frac{m_{\phi }|\tau |}{2},  \notag \\
a(t) &=&\mathrm{e}^{m_{\phi }t/2}\left( 1+\alpha _{0}^{l}(\alpha _{0}m_{\phi
}t)^{n}\right) ,\qquad \frac{m_{\phi }|\tau |}{2}=\mathrm{e}^{-m_{\phi
}t/2}\left( 1+\alpha _{0}^{l}(\alpha _{0}m_{\phi }t)^{n}\right) ,
\label{expas}
\end{eqnarray}%
where products such as $\alpha _{0}^{p}(\alpha _{0}m_{\phi }t)^{q}$ are
symbolic and it is understood that $p+q>0$. It is always possible to switch
from an expansion in $t$ to an expansion in $\ln \left\vert \tau \right\vert 
$.

The formulas just written\ allow us to express everything in terms of
exponentials and powers of $t$. The exponential behaviors, which can be
retrieved from the de Sitter limit, guide us through the superhorizon limit
of the spectra.

Let us recall that the expansion in powers of $\alpha $ is, indeed, the
expansion around the de Sitter limit. Once a contribution is exponentially
suppressed in the superhorizon limit ($t\rightarrow +\infty $) of the de
Sitter limit ($\alpha \rightarrow 0$), it cannot be resuscitated by turning
on the expansion in powers of $\alpha $, since the corrections are just
powers of $t$, as shown above. The following arguments illustrate these
facts more explicitly.

The equations of motion of (\ref{lut}) are%
\begin{equation}
\frac{\mathrm{d}}{\mathrm{d}t}\left( a^{3}\dot{u}_{\mathbf{k}}\right)
+k^{2}au_{\mathbf{k}}=0.  \label{det}
\end{equation}%
Let us first consider the de Sitter limit, where $a(t)=\mathrm{e}^{m_{\phi
}t/2}$. In the superhorizon limit $t\rightarrow +\infty $ ($\tau \rightarrow
0^{-}$), the second term on the left-hand side of (\ref{det}) can be
dropped, so we just have $a^{3}\dot{u}_{\mathbf{k}}=$ constant, which is
solved by%
\begin{equation}
u_{\mathbf{k}}=c\int^{t}\frac{\mathrm{d}t^{\prime }}{a(t^{\prime })^{3}}%
+d=c^{\prime }\mathrm{e}^{-3m_{\phi }t/2}+d^{\prime },  \label{rgu}
\end{equation}%
where $c$, $c^{\prime }$, $d$ and $d^{\prime }$ are integration constants.
From this solution, we see that in the superhorizon limit we can also drop
the contribution $c^{\prime }\mathrm{e}^{-3m_{\phi }t/2}$, since it is
negligible with respect to the constant $d^{\prime }$. We thus obtain $u_{%
\mathbf{k}}=$ constant.

The conclusion holds throughout the RG flow, as we can prove by turning on $%
\alpha $ to move away from the de Sitter limit. Indeed, if we use the
expansions given above, we see that the exponential appearing in (\ref{rgu})
gets multiplied by powers of $t$, which cannot overcome the exponential
factor. Instead, the constant $d^{\prime }$ is unaffected, since the
correction $k^{2}au_{\mathbf{k}}$ appearing in (\ref{det}) is exponentially
subleading, which ensures that $u_{\mathbf{k}}=$ constant solves (\ref{det})
in the superhorizon limit even for nonzero $\alpha $.

We conclude that $u$ tends to a constant for $k|\tau |\rightarrow 0$. By
formulas (\ref{pu}) and (\ref{pt0}), so does $k^{-3}\mathcal{P}_{T}$. This
gives the renormalization-group equation%
\begin{equation}
\frac{\mathrm{d}\mathcal{P}_{T}}{\mathrm{d\ln }|\tau |}=0.  \label{CStens}
\end{equation}%
By equation (\ref{weq}) and the Bunch-Davies condition (\ref{bunch}), the
solution $w$ depends on $\tau $ only through $\eta $ and $\alpha (-\tau )$.
By (\ref{w}), (\ref{pu}) and (\ref{pt0}), so does the spectrum $\mathcal{P}%
_{T}$. In this respect, note that the factor $a$ of (\ref{w}) conspires with
the factor $1/\eta $ of (\ref{superhor}) to give $a\eta =(-aH\tau )k/H$,
which is a power series in $\alpha (-\tau )$ by formulas (\ref{acca}) and (%
\ref{aHtau}). Thus, equation (\ref{CStens}) can be rewritten as the RG
evolution equation 
\begin{equation}
\left( \frac{\partial }{\partial \mathrm{\ln }|\tau |}+\beta _{\alpha }\frac{%
\partial }{\partial \alpha }\right) \mathcal{P}_{T}=0.  \label{RGtensor}
\end{equation}%
The $\alpha $ on the right-hand side of this equation stands for $\alpha
(-\tau )$. Equation (\ref{RGtensor}) has the standard form of the
Callan-Symanzik equation of quantum field theory, with zero anomalous
dimension.

The RG equation tells us that $\mathcal{P}_{T}$ is actually $\tau $
independent, so in the end it is just a function of $\alpha _{0}=\alpha
(1/k) $. Formula (\ref{pt}) provides an explicit check of the property and
so does formula (\ref{prstaro}) below.

Summarizing, the superhorizon limit $k|\tau |\rightarrow 0$ kills the powers
of $k|\tau |$ and the RG evolution equation completes the job by killing the
logarithms of $k|\tau |$, so that, in the end, no dependence on $\tau $
survives.

\subsection{Scalar fluctuations}

\label{scalarostaro}

Now we study the scalar fluctuations in the $R+R^{2}$ theory. We choose the
comoving gauge, where the fluctuation $\delta \phi $ of the scalar field $%
\phi $ is identically zero. The metric is parametrized as%
\begin{equation}
g_{\mu \nu }=\text{diag}(1,-a^{2},-a^{2},-a^{2})+2\text{diag}(\Phi
,a^{2}\Psi ,a^{2}\Psi ,a^{2}\Psi )-\delta _{\mu }^{0}\delta _{\nu
}^{i}\partial _{i}B-\delta _{\mu }^{i}\delta _{\nu }^{0}\partial _{i}B.
\label{mets}
\end{equation}

After Fourier transforming the space coordinates, (\ref{sqgeq}) gives the
quadratic Lagrangian 
\begin{equation}
(8\pi G)\frac{\mathcal{L}_{\text{s}}}{a^{3}}=-3(\dot{\Psi}+H\Phi )^{2}+4\pi G%
\dot{\phi}^{2}\Phi ^{2}+\frac{k^{2}}{a^{2}}\left[ 2B(\dot{\Psi}+H\Phi )+\Psi
(\Psi -2\Phi )\right] ,  \notag
\end{equation}%
omitting the subscripts $\mathbf{k}$ and $-\mathbf{k}$.

Integrating $B$ out, we obtain $\Phi =-\dot{\Psi}/H$. Inserting this
solution back into the action, we find 
\begin{equation}
(8\pi G)\frac{\mathcal{L}_{\text{s}}}{a^{3}}=3\alpha ^{2}\left( \dot{\Psi}%
^{2}-\frac{k^{2}}{a^{2}}\Psi ^{2}\right) .  \label{ls}
\end{equation}%
The equation of motion becomes in the superhorizon limit%
\begin{equation*}
\frac{\mathrm{d}}{\mathrm{d}t}\left( a^{3}\alpha ^{2}\dot{\Psi}\right) =0,
\end{equation*}%
which is solved by%
\begin{equation*}
\dot{\Psi}=\frac{c}{a^{3}\alpha ^{2}},
\end{equation*}%
where $c$ is a constant. Using the expansions (\ref{expas}), it is easy to
see that (in the superhorizon limit) $\dot{\Psi}$ tends exponentially to
zero, so $\Psi $ tends exponentially to a constant. The corrections
proportional to $k^{2}/a^{2}$ in (\ref{ls}) are also exponentially
suppressed. Recalling that $\Psi $ coincides with the curvature perturbation 
$\mathcal{R}$ in the gauge we are using, we obtain the RG equation%
\begin{equation}
\frac{\mathrm{d}\mathcal{P}_{\mathcal{R}}}{\mathrm{d\ln }|\tau |}=0
\label{RGsstaro}
\end{equation}%
for the spectrum $\mathcal{P}_{\mathcal{R}}$ of the $\mathcal{R}$
fluctuations. Again, since the solution $w$ of the Mukhanov-Sasaki equation
depends on $\tau $ only through $\eta $ and $\alpha (-\tau )$, and so does
the spectrum $\mathcal{P}_{\mathcal{R}}$, equation (\ref{RGsstaro}) can be
rewritten as%
\begin{equation}
\left( \frac{\partial }{\partial \mathrm{\ln }|\tau |}+\beta _{\alpha }\frac{%
\partial }{\partial \alpha }\right) \mathcal{P}_{\mathcal{R}}=0.  \label{RGS}
\end{equation}%
The calculations we are going to perform will provide nontrivial checks of
this formula. Like $\mathcal{P}_{T}$, the spectrum $\mathcal{P}_{\mathcal{R}%
} $ has a vanishing anomalous dimension.

Defining%
\begin{equation}
w=\alpha a\Psi \sqrt{\frac{3k}{4\pi G}},  \label{wsstaro}
\end{equation}%
the $w$ action reads%
\begin{equation}
S_{\text{s}}=\frac{1}{2}\int \mathrm{d}\eta \left[ w^{\prime 2}-w^{2}+\left(
\nu _{\text{s}}^{2}-\frac{1}{4}\right) \frac{w^{2}}{\eta ^{2}}\right] ,
\label{muksstaro}
\end{equation}%
where 
\begin{equation*}
\nu _{\text{s}}^{2}-\frac{1}{4}=\left( \frac{\beta _{\alpha }}{\alpha }%
+aH\tau -1+\beta _{\alpha }\frac{\mathrm{d}}{\mathrm{d}\alpha }\right)
\left( \frac{\beta _{\alpha }}{\alpha }+aH\tau \right) .
\end{equation*}%
Using (\ref{aHtau}), we find%
\begin{equation}
\nu _{\text{s}}=\frac{3}{2}+2\alpha +6\alpha ^{2}+\frac{208}{9}\alpha ^{3}+%
\frac{1361}{9}\alpha ^{4}+\mathcal{O}(\alpha ^{5}).  \label{nusstaro}
\end{equation}%
As before, the $w$ equation%
\begin{equation}
w^{\prime \prime }+w-\left( \nu _{\text{s}}^{2}-\frac{1}{4}\right) \frac{w}{%
\eta ^{2}}=0  \label{veqsstaro}
\end{equation}%
can be solved by expanding in powers of $\alpha $, but this time we have to
include a term proportional to $\alpha _{0}$: 
\begin{equation}
w(\eta )=w_{0}(\eta )+\alpha _{0}w_{1}(\eta )+\alpha _{0}^{2}w_{2}(\eta
)+\alpha _{0}^{3}w_{3}(\eta )+\cdots .  \label{wetasstaro}
\end{equation}%
We obtain equations of the form%
\begin{equation}
w_{n}^{\prime \prime }+w_{n}-\frac{2w_{n}}{\eta ^{2}}=\frac{g_{n}(\eta )}{%
\eta ^{4}},  \label{weqsstaro}
\end{equation}%
where the functions $g_{n}$ are determined recursively from $w_{j}$, $j<n$,
as explained in subsection \ref{tensor}. We find $g_{0}(\eta )=0$ and 
\begin{equation}
g_{1}=6w_{0},\qquad g_{2}=2(11-6\ln \eta )w_{0}+g_{1}^{+},\qquad
g_{3}=\left( \frac{280}{3}-98\ln \eta +24\ln ^{2}\eta \right)
w_{0}+g_{2}^{+},  \label{gplus}
\end{equation}%
etc., where $g_{k}^{+}$ means $g_{k}$ with every $w_{m}$ replaced by $%
w_{m+1} $.

The function $w_{0}$ is still (\ref{w0}), while $w_{1}$ coincides with the $%
w_{2}$ of (\ref{w23}) apart from the overall factor:%
\begin{equation}
w_{1}(\eta )=\frac{\sqrt{2}}{\eta }\left[ 2i\mathrm{e}^{i\eta }+(\eta -i)%
\mathrm{e}^{-i\eta }\left( \hspace{0.01in}\text{Ei}(2i\eta )-i\pi \right) %
\right] .  \label{w1s}
\end{equation}%
The functions $w_{k}$ with $k>1$ can be written by quadratures.

Now we study the power spectrum. In the limit $\eta \rightarrow \infty $,
the functions $g_{n}$ and $w_{n}$, $n>0$, satisfy (\ref{inf}) for every $%
\delta >0$. In the superhorizon limit $\eta \rightarrow 0$, we have%
\begin{equation}
g_{n}(\eta )\underset{\eta \sim 0}{\sim }\frac{1}{\eta }P_{n-1}^{(g)}(\ln
\eta )+\mathcal{O}(\ln ^{n-1}\eta )\qquad w_{n}(\eta )\underset{\eta \sim 0}{%
\sim }\frac{1}{\eta }P_{n}^{(w)}(\ln \eta )+\mathcal{O}(\ln ^{n}\eta ).
\label{anda}
\end{equation}%
As in the case of the tensor fluctuations, we can drop the term $w_{n}$ of
equation (\ref{gn}) for $\eta \rightarrow 0$ and the solution (\ref{wn})
simplifies to the form (\ref{appros}).

To the leading log order, it is enough to truncate (\ref{nusstaro}) to order 
$\alpha $ and replace $\alpha $ with (\ref{lead}). Equation (\ref{weqsstaro}%
) then simplifies to%
\begin{equation*}
w^{\prime \prime }+w-\frac{2w}{\eta ^{2}}=\frac{6\alpha _{0}w}{\eta
^{2}(1+2\alpha _{0}\ln \eta )}.
\end{equation*}%
Formulas (\ref{anda}) suggest to parametrize the leading behavior of $w_{n}$
as 
\begin{equation}
w_{n}(\eta )\underset{\eta \sim 0}{\sim }\frac{id_{n}}{\eta \sqrt{2}}(-2\ln
\eta )^{n},  \label{wnas}
\end{equation}%
where $d_{n}$ are constants. Expanding in powers of $\alpha _{0}$, we find
equation (\ref{weqsstaro}) with 
\begin{equation}
g_{n}(\eta )=\frac{3i\sqrt{2}}{\eta }(-2\ln \eta
)^{n-1}\sum_{k=0}^{n-1}d_{k}.  \label{cns}
\end{equation}%
Using (\ref{appros}), we obtain 
\begin{equation}
w_{n}(\eta )\underset{\eta \sim 0}{\sim }\frac{i}{\eta \sqrt{2}}(-2\ln \eta
)^{n}\frac{1}{n}\sum_{k=0}^{n-1}d_{k}\text{.}  \label{wcns}
\end{equation}%
Matching (\ref{wcns}) with (\ref{wnas}), we obtain a recursion relation for
the constants $d_{k}$. Since $d_{0}=1$, all the $d_{k}$ turn out to be equal
to one.

Resumming (\ref{wetasstaro}) we find the leading log correction to $w(\eta )$%
, which is%
\begin{equation}
w(\eta )\underset{\eta \sim 0}{\sim }\frac{i}{\eta \sqrt{2}}\frac{1}{%
1+2\alpha _{0}\ln \eta }.  \label{agree}
\end{equation}%
Now, using (\ref{wsstaro}), the leading log behavior of the curvature
perturbation turns out to be 
\begin{equation}
\mathcal{R}=\Psi =\frac{w}{a\alpha }\sqrt{\frac{4\pi G}{3k}}\sim \sqrt{\frac{%
4\pi G}{3k}}\frac{i}{a\alpha \eta \sqrt{2}}\frac{1}{1+2\alpha _{0}\ln \eta }.
\label{rr}
\end{equation}

The $\mathcal{R}$ power spectrum is defined by%
\begin{equation}
\langle \mathcal{R}_{\mathbf{k}}(\tau )\mathcal{R}_{\mathbf{k}^{\prime
}}(\tau )\rangle =(2\pi )^{3}\delta ^{(3)}(\mathbf{k}+\mathbf{k}^{\prime })%
\frac{2\pi ^{2}}{k^{3}}\mathcal{P}_{\mathcal{R}},\qquad \frac{2\pi ^{2}}{%
k^{3}}\mathcal{P}_{\mathcal{R}}=|\Psi |^{2},  \label{PR}
\end{equation}%
so (\ref{rr}) gives%
\begin{equation}
\mathcal{P}_{\mathcal{R}}=\frac{GH^{2}}{3\pi (aH\tau )^{2}}\frac{1}{\alpha
(-\tau )^{2}(1+2\alpha _{0}\ln \eta )^{2}}.  \label{rrstaro}
\end{equation}%
Using (\ref{acca}) and (\ref{aHtau}) with $\alpha (-\tau )$ given by (\ref%
{lead}), we find%
\begin{equation}
\mathcal{P}_{\mathcal{R}}(k)=\frac{Gm_{\phi }^{2}}{12\pi \alpha _{0}^{2}}=%
\frac{m_{\phi }^{2}G}{12\pi \alpha (1/k)^{2}},  \label{prstaro}
\end{equation}%
which does satisfy the RG equation (\ref{RGS}). As expected, the $\tau $
dependence has disappeared.

The spectral index is defined by%
\begin{equation*}
\frac{\mathrm{d}\ln \mathcal{P}_{\mathcal{R}}(k)}{\mathrm{d}\ln k}=n_{%
\mathcal{R}}(k)-1.
\end{equation*}%
We find%
\begin{equation}
n_{\mathcal{R}}-1=-4\alpha ,\qquad \frac{\mathrm{d}n_{\mathcal{R}}}{\mathrm{d%
}\ln k}=4\beta _{\alpha }=-8\alpha ^{2},\qquad \frac{\mathrm{d}^{n}n_{%
\mathcal{R}}}{\mathrm{d}\ln k\hspace{0.01in}^{n}}=-2^{n+2}n!\alpha ^{n+1},
\label{coesstaro}
\end{equation}%
where $\alpha $ stands for $\alpha (1/k)$. Again, the leading log
contributions to the running, the running of the running, etc., are all
related to the leading log contribution to $n_{\mathcal{R}}-1$, and
ultimately $\mathcal{P}_{\mathcal{R}}$, by the RG equation.

\subsection{Subleading log corrections}

Now we study the subleading log corrections to the power spectra of the
tensor and scalar fluctuations. It is convenient to expand the
Mukhanov-Sasaki equation in a more systematic way. We write it as%
\begin{equation}
w^{\prime \prime }+w-2\frac{w}{\eta ^{2}}=\sigma \frac{w}{\eta ^{2}},
\label{mukhagen}
\end{equation}%
where $\sigma =\nu ^{2}-(9/4)=\mathcal{O}(\alpha )$ is a power series in $%
\alpha $, while $\nu $ is $\nu _{\text{t}}$ or $\nu _{\text{s}}$.

We start by studying some general properties of this equation. Following (%
\ref{superhor}) and (\ref{anda}), we separate $\eta w(\eta )$ into a power
series $Q(\ln \eta )$ in $\ln \eta $ plus the rest:%
\begin{equation}
\eta w=Q(\ln \eta )+W(\eta ),  \label{decompo}
\end{equation}%
where $W(\eta )$ is an expansion in powers of $\eta $ and logarithms $\ln
\eta $, such that $W(\eta )\rightarrow 0$ term-by-term for $\eta \rightarrow
0$. Then we observe that if the right-hand side of (\ref{mukhagen}) is%
\begin{equation}
\sigma \frac{w}{\eta ^{2}}=\frac{c}{\eta ^{3-m}}\ln ^{n}\eta ,  \label{smn}
\end{equation}%
where $c$ is a constant and $n,m\geqslant 0$ are integers, it is easy to
show, using (\ref{wn}) or (\ref{appros}), that only $m=0$ contributes to $%
Q(\ln \eta )$. Such contributions are equal to%
\begin{equation}
Q(\ln \eta )=-\frac{c}{9}\sum_{k=-1}^{n-1}3^{-k}\frac{\mathrm{d}^{k}}{%
\mathrm{d}\ln ^{k}\eta }\ln ^{n}\eta +\text{ constant},  \label{qlog}
\end{equation}%
where it is understood that the \textquotedblleft $-1$ derivative%
\textquotedblright\ ($k=0$) is the integral from 0 to $\ln \eta $:%
\begin{equation*}
\frac{\mathrm{d}^{-1}f(x)}{\mathrm{d}x^{-1}}\equiv \int_{0}^{x}f\left(
x^{\prime }\right) \hspace{0.01in}\mathrm{d}x^{\prime }.
\end{equation*}%
The constant in (\ref{qlog}) is left unspecified, because it does not
provide useful information at this level, for the reasons we explain below.
Summing the contributions due to (\ref{smn}) for $m=0$, $n\geqslant 0$, we
find the equation satisfied by $Q$, which is 
\begin{equation}
Q=-\frac{1}{9}\sum_{n=-1}^{\infty }3^{-n}\frac{\mathrm{d}^{n}(\sigma Q)}{%
\mathrm{d}\ln ^{n}\eta }+\text{ constant}.  \label{PWeq}
\end{equation}%
Differentiating it once, we get 
\begin{equation}
\frac{\mathrm{d}Q}{\mathrm{d}\ln \eta }=-\frac{\sigma }{3}Q-\frac{1}{3}%
\sum_{n=1}^{\infty }3^{-n}\frac{\mathrm{d}^{n}(\sigma Q)}{\mathrm{d}\ln
^{n}\eta }.  \label{Pw}
\end{equation}

As for the leading logs, it is sufficient to truncate this equation to the
first term on the right-hand side. Then the equation integrates to%
\begin{equation}
Q(\ln \eta )=Q(0)\exp \left( -\frac{1}{3}\int_{0}^{\ln \eta }\sigma (\alpha
(\eta ^{\prime }/k))\hspace{0.01in}\mathrm{d}\ln \eta ^{\prime }\right)
=Q(0)\exp \left( -\frac{1}{3}\int_{\alpha _{0}}^{\alpha (-\tau )}\frac{%
\sigma (\alpha ^{\prime })}{\beta _{\alpha }(\alpha ^{\prime })}\hspace{%
0.01in}\mathrm{d}\alpha ^{\prime }\right) .  \label{leadingPw}
\end{equation}

To verify this formula in the cases of the tensor and scalar fluctuations,
it is sufficient to use formula (\ref{lead}) for the running coupling and
truncate the beta function to its first contribution, which is $-2\alpha
^{2} $. In the tensor case, formula (\ref{nt}) gives $\sigma =9\alpha ^{2}$
to the lowest order, so we obtain%
\begin{equation*}
Q(\ln \eta )=Q(0)\exp \left( -\frac{3\alpha _{0}^{2}\ln \eta }{1+2\alpha
_{0}\ln \eta }\hspace{0.01in}\right) =Q(0)\left( 1-\frac{3\alpha _{0}^{2}\ln
\eta }{1+2\alpha _{0}\ln \eta }\hspace{0.01in}+\text{ subleading}\right) ,
\end{equation*}%
which agrees with (\ref{w0tensor}) for $Q(0)=i/\sqrt{2}$. In the scalar
case, formula (\ref{nusstaro}) gives $\sigma =6\alpha $ to the lowest order,
so we find%
\begin{equation*}
Q(\ln \eta )=\frac{Q(0)}{1+2\alpha _{0}\ln \eta },
\end{equation*}%
which agrees with (\ref{agree}) for $Q(0)=i/\sqrt{2}$ again.

The integration constant $Q(0)$ cannot be computed just from the behaviors
of the functions for $\eta \rightarrow 0$. Indeed, it is not enough to know
the integrand of (\ref{wn}) for $\eta \sim 0$ to calculate the contributions
to the integral that have the form constant$/\eta $. Yet, equation (\ref{Pw}%
) is enough to perform a full check of the RG equations (\ref{RGtensor}) and
(\ref{RGS}), since it encodes all the contributions $\sim (\ln ^{n}\eta
)/\eta $ with $n>0$.

We first verify the validity of the evolution equations to the
next-to-leading log order and later compute $Q(0)$ to the same order. The
next-to-leading log corrections can be studied by keeping one term more on
the right-hand-side of (\ref{Pw}), as well as in the beta function, the
running coupling and the expression of $\sigma $. Specifically, the beta
function can be truncated to $\beta _{\alpha }=-2\alpha ^{2}-(5/3)\alpha
^{3} $. As for the running coupling, we can use formula (\ref{runca}). The
formulas of $\sigma $ can instead be truncated to $\sigma =\nu _{\text{t}%
}^{2}-(9/4)=9\alpha ^{2}+48\alpha ^{3}$ and $\sigma =\nu _{\text{s}%
}^{2}-(9/4)=6\alpha +22\alpha ^{2}$ for tensors and scalars, respectively.

The truncated version of (\ref{Pw}), which is%
\begin{equation*}
\frac{\mathrm{d}Q}{\mathrm{d}\ln \eta }=-\frac{\sigma }{3}Q-\frac{1}{9}\frac{%
\mathrm{d}(\sigma Q)}{\mathrm{d}\ln \eta },
\end{equation*}%
is solved by%
\begin{equation}
Q(\ln \eta )=Q(0)\frac{9+\sigma (\alpha _{0})}{9+\sigma (\alpha (-\tau ))}%
\exp \left( -\int_{\alpha _{0}}^{\alpha (-\tau )}\frac{3\sigma (\alpha
^{\prime })\mathrm{d}\alpha ^{\prime }}{(9+\sigma (\alpha ^{\prime }))\beta
_{\alpha }(\alpha ^{\prime })}\hspace{0.01in}\right) .  \label{nleadingPw}
\end{equation}

Moreover, the power spectra (\ref{pt0}) and (\ref{PR}) require formulas (\ref%
{acca}) and (\ref{aHtau}) up to the order $\alpha ^{2}$, so we use%
\begin{equation*}
H=\frac{m_{\phi }}{2}\left( 1-\frac{3\alpha }{2}+\frac{7\alpha ^{2}}{4}%
\right) ,\qquad -aH\tau =1+3\alpha ^{2}.
\end{equation*}%
Putting all the ingredients together, we find%
\begin{eqnarray}
\mathcal{P}_{T} &=&\frac{8Gm_{\phi }^{2}}{\pi }|Q^{(T)}(0)|^{2}\left(
1-3\alpha (1/k)-\frac{1}{4}\alpha (1/k)^{2}\right) ,  \notag \\
\mathcal{P}_{\mathcal{R}} &=&\frac{Gm_{\phi }^{2}}{6\pi }\frac{|Q^{(\mathcal{%
R)}}(0)|^{2}}{\alpha (1/k)^{2}}\left( 1-3\alpha (1/k)\right) .  \label{ptprQ}
\end{eqnarray}%
for tensors and scalars, respectively. We see that the $\tau $ dependence
has disappeared again, in agreement with the RG equations.

Now we come to the integration constant $Q(0)$. In general, we have%
\begin{equation}
Q(0)=\frac{i}{\sqrt{2}}h(\alpha _{0}),  \label{Q0}
\end{equation}%
where $h(\alpha _{0})=1+\mathcal{O}(\alpha _{0})$ is a power series in $%
\alpha _{0}$. As said, it is not necessary to know $Q(0)$ to check the RG
equation. Yet, its knowledge is crucial to determine the subleading log
corrections to the power spectra. The constants $Q^{(T\mathcal{)}}(0)$ and $%
Q^{(\mathcal{R)}}(0)$ of $\mathcal{P}_{T}$ and $\mathcal{P}_{\mathcal{R}}$
can be calculated from the definition (\ref{decompo}) by means of the
complete expressions of the functions $w_{n}$. Using formulas (\ref{w23})
and (\ref{w1s}), we find%
\begin{eqnarray}
Q^{(T\mathcal{)}}(\ln \eta ) &=&\frac{i}{\sqrt{2}}\left( 1+\frac{3}{2}\alpha
_{0}^{2}(4-2\gamma _{M}+i\pi -2\ln \eta )\right) +\mathcal{O}(\alpha
_{0}^{3}),  \notag \\
Q^{(\mathcal{R)}}(\ln \eta ) &=&\frac{i}{\sqrt{2}}%
\Big(%
1+\alpha _{0}(4-2\gamma _{M}+i\pi -2\ln \eta )%
\Big)%
+\mathcal{O}(\alpha _{0}^{2}).  \label{qtqr}
\end{eqnarray}%
In the end, we obtain%
\begin{eqnarray}
\mathcal{P}_{T}(k) &=&\frac{4Gm_{\phi }^{2}}{\pi }\left[ 1-3\alpha
(1/k)+\left( \frac{47}{4}-6\gamma _{M}\right) \alpha (1/k)^{2}\right] , 
\notag \\
\mathcal{P}_{\mathcal{R}}(k) &=&\frac{Gm_{\phi }^{2}}{12\pi \alpha (1/k)^{2}}%
\Big(%
1+(5-4\gamma _{M})\alpha (1/k)%
\Big)%
.  \label{ptpr}
\end{eqnarray}%
The results agree with those of ref. \cite{ABP} in the limit $m_{\chi
}\rightarrow \infty $. In addition, now we know how to use them to calculate
the subleading log corrections to the spectral indices and their running
coefficients. For the tensor fluctuations, we find%
\begin{eqnarray}
n_{T} &=&-\beta _{\alpha }(\alpha )\frac{\partial \ln P_{T}}{\partial \alpha 
}=-6\alpha ^{2}+24\alpha ^{3}(1-\gamma _{M}),  \notag \\
\frac{\mathrm{d}n_{T}}{\mathrm{d}\ln k} &=&-\beta _{\alpha }(\alpha )\frac{%
\partial n_{T}}{\partial \alpha }=-24\alpha ^{3}+4\alpha ^{4}(31-36\gamma
_{M}),  \label{runT} \\
\frac{\mathrm{d}^{2}n_{T}}{\mathrm{d}\ln k\hspace{0.01in}\hspace{0.01in}^{2}}
&=&-\beta _{\alpha }(\alpha )\frac{\partial }{\partial \alpha }\left( -\beta
_{\alpha }(\alpha )\frac{\partial n_{T}}{\partial \alpha }\right)
=-144\alpha ^{4}+8\alpha ^{5}(109-144\gamma _{M}).  \notag
\end{eqnarray}%
etc., where now $\alpha $ stands for $\alpha (1/k)$. For the scalar
fluctuations, we find%
\begin{eqnarray}
n_{\mathcal{R}}-1 &=&-4\alpha +\frac{4}{3}\alpha ^{2}(5-6\gamma _{M}), 
\notag \\
\frac{\mathrm{d}n_{R}}{\mathrm{d}\ln k} &=&-8\alpha ^{2}+4\alpha ^{3}\left(
5-8\gamma _{M}\right) ,\qquad \frac{\mathrm{d}^{2}n_{\mathcal{R}}}{\mathrm{d}%
\ln k\hspace{0.01in}\hspace{0.01in}^{2}}=-32\alpha ^{3}+\frac{8}{3}\alpha
^{4}\left( 35-72\gamma _{M}\right) ,  \label{runS}
\end{eqnarray}%
etc.

The results of this section agree with those available in the literature 
\cite{run1,run2} up to the resummations of the leading logs, which means
once we expand $\alpha (1/k)$ in powers of the \textquotedblleft pivot
coupling\textquotedblright\ $\alpha _{\ast }=\alpha (1/k_{\ast })$, i.e. the
running coupling evaluated at the pivot\ scale $k_{\ast }$, and consider the
product $\alpha _{\ast }\ln (k/k_{\ast })$ of order one instead of order
zero.

\section{Quantum gravity}

\label{qg}\setcounter{equation}{0}

In this section we generalize the results to quantum gravity. We begin by
proving that the RG equations still hold.

\subsection{RG evolution equations of the power spectra}

We prove that the power spectra obey the RG evolution equations (\ref%
{RGtensor}) and (\ref{RGS}), in the superhorizon limit. We refer to the
notation and formulas of ref. \cite{ABP}.

In the case of the tensor fluctuations, we take the parametrization (\ref%
{met}) of the metric and expand the action (\ref{sqgeq}). The quadratic
Lagrangian in the superhorizon limit reads 
\begin{equation}
\mathcal{L}_{\text{t}}=\frac{a^{3}}{8\pi G}\left[ 1+\frac{2H^{2}}{m_{\chi
}^{2}}\left( 1-\frac{3}{2}\alpha ^{2}\right) \right] \dot{u}^{2}-\frac{\ddot{%
u}^{2}}{m_{\chi }^{2}}.  \label{lthor}
\end{equation}%
We see that the field equations admit the obvious solution $u=$ constant,
since $\mathcal{L}_{\text{t}}$ depends only on the derivatives of $u$. We
want to show that this is the only solution that survives the superhorizon
limit. Around the de Sitter space we can use the formulas (\ref{expas}),
which expand the solution in terms of exponentials multiplied by power
series. The exponential factors can be derived from the de Sitter limit,
where the Lagrangian becomes 
\begin{equation*}
\mathcal{L}_{\text{t}}=\frac{\mathrm{e}^{3m_{\phi }t/2}}{8\pi G}\left[ \dot{u%
}^{2}\left( 1+\frac{m_{\phi }^{2}}{2m_{\chi }^{2}}\right) -\frac{\ddot{u}^{2}%
}{m_{\chi }^{2}}\right] .
\end{equation*}%
The most general solution of its equation of motion is%
\begin{equation}
u(t)=c_{1}+c_{2}\mathrm{e}^{-3m_{\phi }t/2}+\mathrm{e}^{-3m_{\phi
}t/4}\left( c_{3}\mathrm{e}^{ist}+c_{4}\mathrm{e}^{-ist}\right) ,  \label{ut}
\end{equation}%
where $s=\sqrt{m_{\chi }^{2}-(m_{\phi }^{2}/16)}$, which is real by the
consistency condition $m_{\chi }>m_{\phi }/4$ of the fakeon projection \cite%
{ABP}. We see that in the superhorizon limit $t\rightarrow +\infty $ only
the constant $c_{1}$ survives. The conclusion does not change by moving away
from the de Sitter limit, since the exponentials of (\ref{ut}) just get
multiplied by powers of $t$.

The arguments just given apply to the unprojected action (\ref{sqgeq}) and
the whole space of solutions of its field equations. Since the relevant
solution $u=$ constant belongs to the physical subspace, the conclusions
extend to the projected action as well. Details on the fakeon projection are
given below. Ultimately, by (\ref{pu}) and (\ref{pt0}) the tensor power
spectrum $P_{T}$ satisfies equation (\ref{CStens}), from which (\ref%
{RGtensor}) follows.

In the case of the scalar fluctuations, we expand the action (\ref{sqgeq})
to the quadratic order with the metric (\ref{mets}) (for the result of this
operation, see \cite{ABP}). Next, we remove $\Phi $, which is an auxiliary
field, by means of its own field equation. Third, we define%
\begin{equation}
\Psi =U+f(t)B,  \label{psub}
\end{equation}%
where $f(t)$ is a function, and express the Lagrangian in terms of $B$ and
the new field $U$. We choose $f$ to remove the term proportional to $\dot{B}%
^{2}$. After integrating by parts to eliminate $\dot{B}$ altogether, $B$
turns into an auxiliary field as well, so we remove it by means of its own
field equation. Then we take the superhorizon limit $k/(aH)\rightarrow 0$
and see that $\mathcal{L}_{\text{s}}$ and $B$ depend only on the derivatives
of $U$, but not on $U$ itself, so $\Psi =$ constant is a solution.
Explicitly, if we expand around the de Sitter metric, we obtain 
\begin{equation}
(8\pi G)\frac{\mathcal{L}_{\text{s}}}{a^{3}}=3\alpha ^{2}\left( \dot{U}^{2}-%
\frac{2\ddot{U}^{2}}{2m_{\chi }^{2}+m_{\phi }^{2}}\right) ,\qquad \Psi =U+%
\frac{3m_{\phi }\dot{U}+2\ddot{U}}{2m_{\chi }^{2}+m_{\phi }^{2}},  \notag
\end{equation}%
up to higher powers of $\alpha $. The expression of $\Psi $ is obtained from
(\ref{psub}) after substituting $B$ with its solution. Using (\ref{expas})
again, the solution of the $U$ field equation has the form (\ref{ut}) up to
powers of $t$ multiplying the exponential factors, so $\Psi $ tends to a
constant in the superhorizon limit. The conclusion does not change moving
away from the de Sitter fixed point. Moreover, it applies to the unprojected
action and extends immediately to the projected one. This proves that the
scalar power spectrum satisfies the RG evolution equations (\ref{RGsstaro})
and (\ref{RGS}).

\subsection{Tensor fluctuations}

\label{tensorQG}

Parametrizing the metric as (\ref{met}), the quadratic Lagrangian obtained
from (\ref{sqgeq}) is 
\begin{equation}
(8\pi G)\frac{\mathcal{L}_{\text{t}}}{a^{3}}=\dot{u}^{2}-\frac{k^{2}}{a^{2}}%
u^{2}-\frac{1}{m_{\chi }^{2}}\left[ \ddot{u}^{2}-2\left( H^{2}-\frac{3}{2}%
\alpha ^{2}H^{2}+\frac{k^{2}}{a^{2}}\right) \dot{u}^{2}+\frac{k^{4}}{a^{4}}%
u^{2}\right] ,  \label{lt}
\end{equation}%
plus an identical contribution for $v$. We can eliminate the higher
derivatives by means of the procedure used in \cite{ABP}. Specifically, we
add an auxiliary field $U$ and consider the extended Lagrangian%
\begin{equation}
\mathcal{L}_{\text{t}}^{\prime }=\mathcal{L}_{\text{t}}+\Delta \mathcal{L}_{%
\text{t}},  \label{ltp}
\end{equation}%
where 
\begin{eqnarray}
(8\pi Gm_{\chi }^{2})\frac{\Delta \mathcal{L}_{\text{t}}}{a^{3}} &=&\left[
m_{\chi }^{2}\gamma U-\ddot{u}-\left( 3H-\frac{12\alpha ^{2}H^{3}}{m_{\chi
}^{2}\gamma }\right) \dot{u}\right.  \notag \\
&&\left. -\left( m_{\chi }^{2}\gamma +\frac{k^{2}}{a^{2}}+\frac{3\alpha
^{2}H^{2}(m_{\chi }^{2}-4H^{2})}{m_{\chi }^{2}\gamma }+\frac{24\alpha
^{3}H^{4}}{m_{\chi }^{2}\gamma }\right) u\right] ^{2}  \label{dlt}
\end{eqnarray}%
and 
\begin{equation}
\gamma =1+2\frac{H^{2}}{m_{\chi }^{2}}.  \label{gam}
\end{equation}%
It is immediate to show that $\mathcal{L}_{\text{t}}^{\prime }$ is
equivalent to $\mathcal{L}_{\text{t}}$ by replacing $U$, which appears
algebraically, with the solution of its own field equation. Note that we
have kept an additional term (the one of order $\alpha ^{3}$) with respect
to the formulas of \cite{ABP}. The reason is that, if we want to check the
RG evolution equation to the order $\alpha ^{2}$, it is necessary to make
calculations to the order $\alpha ^{3}$ included.

We diagonalize $\mathcal{L}_{\text{t}}^{\prime }$ in the de Sitter limit by
introducing a second field $V$ such that 
\begin{equation}
u=U+V.  \label{uU}
\end{equation}%
To the order we need, the Lagrangian $\mathcal{L}_{\text{t}}^{\prime }$
takes the form%
\begin{eqnarray}
(8\pi G)\frac{\mathcal{L}_{\text{t}}^{\prime }}{a^{3}\gamma } &=&\dot{U}^{2}-%
\frac{k^{2}}{a^{2}}\left( 1-\frac{12\alpha ^{2}H^{2}}{m_{\chi }^{2}\gamma
^{2}}-\frac{96\alpha ^{3}H^{4}}{m_{\chi }^{4}\gamma ^{2}}\right) U^{2}+\frac{%
36\alpha ^{3}H^{4}}{m_{\chi }^{4}\gamma ^{2}}(m_{\chi }^{2}-4H^{2})U^{2} 
\notag \\
&&-\dot{V}^{2}+\left( m_{\chi }^{2}\gamma +\frac{k^{2}}{a^{2}}\right) V^{2}+%
\frac{6\alpha ^{2}}{m_{\chi }^{2}\gamma }H^{2}\left( m_{\chi }^{2}-4H^{2}+%
\frac{4k^{2}}{\gamma a^{2}}\right) UV.  \label{ltg}
\end{eqnarray}

The fakeon projection amounts to remove the field $V$ by replacing it with a
special solution of its own field equations, which is defined by the fakeon
Green function derived in \cite{ABP}. For the moment, it is sufficient to
know that the solution exists, because from (\ref{ltg}) it is evident that
the projection equates $V$ to something of order $\alpha ^{2}$. Once the
solution is inserted back into (\ref{ltg}), the second line turns out to be $%
\mathcal{O}(\alpha ^{4})$. Therefore, the projected action is given by the
first line of (\ref{ltg}) to the order $\alpha ^{3}$ included. We will need $%
V$ later on, though, since it enters formula (\ref{uU}).

Defining%
\begin{eqnarray}
w &=&\frac{a\sqrt{k\gamma }}{\sqrt{4\pi G}}U,\qquad \sigma =\frac{18m_{\chi
}^{2}\alpha ^{2}}{m_{\phi }^{2}+2m_{\chi }^{2}}+\frac{6m_{\chi }^{2}\alpha
^{3}(32m_{\chi }^{2}+43m_{\phi }^{2})}{(m_{\phi }^{2}+2m_{\chi }^{2})^{2}}+%
\mathcal{O}(\alpha ^{4}),  \notag \\
h &=&1-\frac{12m_{\chi }^{2}m_{\phi }^{2}\alpha ^{2}}{(m_{\phi
}^{2}+2m_{\chi }^{2})^{2}}-\frac{12m_{\phi }^{2}\alpha ^{3}(2m_{\phi
}^{4}+7m_{\chi }^{2}m_{\phi }^{2}-6m_{\chi }^{4})}{(m_{\phi }^{2}+2m_{\chi
}^{2})^{3}}+\mathcal{O}(\alpha ^{4}),  \label{ww}
\end{eqnarray}%
then using (\ref{acca}) and (\ref{aHtau}) and switching to the conformal
time (\ref{tau}), the projected $w$ action to order $\alpha ^{3}$ reads%
\begin{equation}
S_{\text{t}}^{\text{prj}}=\frac{1}{2}\int \mathrm{d}\eta \left( w^{\prime
2}-hw^{2}+2\frac{w^{2}}{\eta ^{2}}+\sigma \frac{w^{2}}{\eta ^{2}}\right) ,
\label{sredF}
\end{equation}%
where the prime denotes the derivative with respect to $\eta =-k\tau $. Due
to the function $h$ in front of $w^{2}$, this action is not of the form (\ref%
{sred}) and its equation of motion is not of the form (\ref{mukhagen}).
Although the term $-hw^{2}$ is negligible in the superhorizon limit, it is
important in the opposite limit, through the Bunch-Davies vacuum condition.

The Bunch-Davies condition is necessary to calculate the constants $Q^{(T%
\mathcal{)}}(0)$ and $Q^{(\mathcal{R)}}(0)$ of formulas (\ref{ptprQ}), but
it is unnecessary to check RG invariance. For these reasons, we first check
the RG evolution equation to the order $\alpha ^{2}$, where we can ignore $%
-hw^{2}$, and then compute the constants $Q(0)$.

The first formula of (\ref{ww}) relates $w$ to $U$, but formula (\ref{uU})
tells us that to work out the $u$ spectrum we also need the relation between 
$V$ and $U$. Such a relation is provided by the fakeon projection, which can
be borrowed to the order we need from \cite{ABP}. Here we just recall the
basic steps that lead to the result.

The $V$ equation of motion derived from (\ref{ltg}) is%
\begin{equation}
\left( \Sigma _{0}+m_{\chi }^{2}+\frac{m_{\phi }^{2}}{2}\right) V=-\frac{%
3\alpha ^{2}m_{\phi }^{2}}{2(m_{\phi }^{2}+2m_{\chi }^{2})}\left( m_{\chi
}^{2}-m_{\phi }^{2}+\frac{8m_{\chi }^{2}k^{2}}{(m_{\phi }^{2}+2m_{\chi
}^{2})a^{2}}\right) U,  \label{Veq}
\end{equation}%
where%
\begin{equation*}
\Sigma _{0}\equiv \frac{\mathrm{d}^{2}}{\mathrm{d}t^{2}}+\frac{3m_{\phi }}{2}%
\frac{\mathrm{d}}{\mathrm{d}t}+\frac{k^{2}}{a^{2}},
\end{equation*}%
We can easily solve (\ref{Veq}) in the superhorizon limit up to corrections
of higher orders in $\alpha $. In that limit, we can ignore the terms
proportional to $k^{2}/a^{2}$. Moreover, (\ref{equa}) gives $[\Sigma
_{0},\alpha ]=\mathcal{O}(\alpha ^{2})$ and the $U$ equation of motion gives 
$\Sigma _{0}U=\mathcal{O}(\alpha )$. Basically, we can commute $\Sigma _{0}$
with $\alpha ^{2}$ and drop $\Sigma _{0}U$, which leads to 
\begin{equation}
V=-\frac{3m_{\phi }^{2}\left( m_{\chi }^{2}-m_{\phi }^{2}\right) }{m_{\phi
}^{2}+2m_{\chi }^{2}}\left. \frac{1}{2\Sigma _{0}+2m_{\chi }^{2}+m_{\phi
}^{2}}\right\vert _{\text{f}}\alpha ^{2}U=-\frac{3\alpha ^{2}m_{\phi
}^{2}(m_{\chi }^{2}-m_{\phi }^{2})}{(m_{\phi }^{2}+2m_{\chi }^{2})^{2}}U,
\label{v}
\end{equation}%
where the subscript f denotes the fakeon prescription. Its role is to ensure
that the solution is the one written on the right-hand side, with no
additions proportional to the solutions of the homogeneous equation.

At this point, we have all the ingredients we need to use formula (\ref%
{nleadingPw}), which gives%
\begin{equation}
\mathcal{P}_{T}=\frac{16Gm_{\chi }^{2}m_{\phi }^{2}}{\pi (m_{\phi
}^{2}+2m_{\chi }^{2})}|Q^{(T)}(0)|^{2}\left[ 1-\frac{6m_{\chi }^{2}\alpha
(1/k)}{m_{\phi }^{2}+2m_{\chi }^{2}}-\alpha (1/k)^{2}\frac{m_{\chi
}^{2}(73m_{\phi }^{2}+2m_{\chi }^{2})}{2(m_{\phi }^{2}+2m_{\chi }^{2})^{2}}%
\right] .  \label{ptQ}
\end{equation}%
As expected, the $\tau $ dependence disappears completely, in agreement with
the RG evolution equation (\ref{CStens}).

The final step is to calculate $Q^{(T)}(0)$, for which it is important to
deal with the term $-hw^{2}$ of (\ref{sredF}). We can rewrite (\ref{sredF})
in the form studied so far by making a change of variables from $\eta $ to $%
\tilde{\eta}(\eta )$, with $\tilde{\eta}^{\prime }(\eta )=\sqrt{h(\eta )}$,
and defining%
\begin{equation}
\tilde{w}(\tilde{\eta}(\eta ))=h(\eta )^{1/4}w(\eta ),\qquad \tilde{\sigma}=%
\frac{\tilde{\eta}^{2}(\sigma +2)}{\eta ^{2}h}+\frac{\tilde{\eta}^{2}}{%
16h^{3}}\left( 4hh^{\prime \prime }-5h^{\prime \hspace{0.01in}2}\right) -2.
\label{wtilde}
\end{equation}%
So doing, (\ref{sredF}) is recast as%
\begin{equation}
\tilde{S}_{\text{t}}^{\text{prj}}=\frac{1}{2}\int \mathrm{d}\tilde{\eta}%
\left( \tilde{w}^{\prime 2}-\tilde{w}^{2}+\frac{2\tilde{w}^{2}}{\tilde{\eta}%
^{2}}+\tilde{\sigma}\frac{\tilde{w}^{2}}{\tilde{\eta}^{2}}\right) ,
\label{s2}
\end{equation}%
where the prime on $\tilde{w}$ denotes the derivative with respect to $%
\tilde{\eta}$.

We have to study the expansion in powers of $\alpha _{0}$. Observe that,
using (\ref{ww}), $\tilde{\sigma}$ turns out to be $\mathcal{O}\left( \alpha
_{0}^{2}\right) $, once we insert the expression\ (\ref{runca}) of the
running coupling $\alpha $. Using the Bunch-Davies condition (\ref{bunch})
in the variable $\tilde{\eta}$, which reads 
\begin{equation}
\tilde{w}(\tilde{\eta})\sim \frac{\mathrm{e}^{i\tilde{\eta}}}{\sqrt{2}}%
\qquad \text{for }\tilde{\eta}\rightarrow \infty ,  \label{BDt}
\end{equation}%
the solution of the $\tilde{S}_{\text{t}}^{\text{prj}}$ equation of motion
has the form 
\begin{equation}
\tilde{w}(\tilde{\eta})=\frac{(\tilde{\eta}+i)\mathrm{e}^{i\tilde{\eta}}}{%
\sqrt{2}\tilde{\eta}}+\alpha _{0}^{2}\Delta \tilde{w}(\tilde{\eta}),\qquad
\lim_{\tilde{\eta}\rightarrow \infty }\Delta \tilde{w}(\tilde{\eta})=0.
\label{wti}
\end{equation}%
Indeed, we know that for $\alpha _{0}=0$ the exact solution is the written
one, by (\ref{BDt}). However, we also know that (\ref{BDt}) must hold for
every $\alpha _{0}$. This means that the corrections neglected in (\ref{wti}%
) must be $\mathcal{O}(\alpha _{0}^{2})$ and disappear in the limit $\tilde{%
\eta}\rightarrow \infty $.

Switching back to the variable $\eta $ by means of the formula%
\begin{equation*}
\tilde{\eta}(\eta )=\int_{0}^{\eta }\sqrt{h(\eta ^{\prime })}d\eta ^{\prime
},
\end{equation*}%
we can work out the vacuum condition for (\ref{sredF}) from (\ref{wti}).
Using the first equation of (\ref{wtilde}), we find 
\begin{equation}
\tilde{\eta}\simeq \eta \left( 1-\frac{6\alpha _{0}^{2}m_{\chi }^{2}m_{\phi
}^{2}}{(m_{\phi }^{2}+2m_{\chi }^{2})^{2}}\right) ,\qquad w(\eta )=\tilde{w}(%
\tilde{\eta}(\eta ))h(\eta )^{-1/4}\simeq \frac{\mathrm{e}^{i\eta }}{\sqrt{2}%
}\left[ 1+\frac{3(3-2i\eta )\alpha _{0}^{2}m_{\chi }^{2}m_{\phi }^{2}}{%
(m_{\phi }^{2}+2m_{\chi }^{2})^{2}}\right] ,  \label{BRtilde}
\end{equation}%
up to terms that are of higher orders in $\alpha _{0}$ or negligible for $%
\eta $ large. If we want to push the calculations to sub-subleading log
orders, other terms of the large $\tilde{\eta}$ expansion in (\ref{wti})
need to be kept.

By expanding $w$ in powers of $\alpha _{0}$ as in (\ref{weta}), we derive
the differential equations obeyed by the functions $w_{n}$. Formula (\ref%
{BRtilde}) gives the asymptotic conditions for the first two, which are%
\begin{equation*}
w_{0}(\eta )\simeq \frac{\mathrm{e}^{i\eta }}{\sqrt{2}},\qquad w_{2}(\eta
)\simeq \frac{3m_{\chi }^{2}m_{\phi }^{2}\mathrm{e}^{i\eta }(3-2i\eta )}{%
\sqrt{2}(m_{\phi }^{2}+2m_{\chi }^{2})^{2}},\qquad \text{for }\eta \text{
large}.
\end{equation*}

We find (\ref{w0}) and%
\begin{equation}
w_{2}(\eta )=\frac{3\sqrt{2}m_{\chi }^{2}}{\eta (m_{\phi }^{2}+2m_{\chi
}^{2})}\left[ 2i\mathrm{e}^{i\eta }\left( 1+\frac{m_{\phi }^{2}(3-3i\eta
-2\eta ^{2})}{4(m_{\phi }^{2}+2m_{\chi }^{2})}\right) +(\eta -i)\mathrm{e}%
^{-i\eta }\left( \hspace{0.01in}\text{Ei}(2i\eta )-i\pi \right) \right] ,
\label{w2F}
\end{equation}%
which upgrades the function $w_{2}$ of formula (\ref{w23}). Studying the $%
\eta \rightarrow 0$ behavior of $w(\eta )$ we can extract $Q(\ln \eta )$ by
means of the decomposition (\ref{decompo}). The outcome leads to 
\begin{equation*}
Q^{(T)}(0)=\frac{i}{\sqrt{2}}\left[ 1+\frac{3\alpha _{0}^{2}m_{\chi
}^{2}(7m_{\phi }^{2}+8m_{\chi }^{2})}{(m_{\phi }^{2}+2m_{\chi }^{2})^{2}}-%
\frac{3\alpha _{0}^{2}m_{\chi }^{2}(2\gamma _{M}-i\pi )}{m_{\phi
}^{2}+2m_{\chi }^{2}}\right] ,
\end{equation*}%
which upgrades the result encoded in the first line of (\ref{qtqr}).
Finally, inserting $Q^{(T)}(0)$ into (\ref{ptQ}) we obtain%
\begin{equation}
\mathcal{P}_{T}=\frac{8Gm_{\chi }^{2}m_{\phi }^{2}}{\pi (m_{\phi
}^{2}+2m_{\chi }^{2})}\left[ 1-\frac{6m_{\chi }^{2}\alpha (1/k)}{m_{\phi
}^{2}+2m_{\chi }^{2}}+\frac{36m_{\chi }^{4}\alpha (1/k)^{2}}{(m_{\phi
}^{2}+2m_{\chi }^{2})^{2}}+\frac{\alpha (1/k)^{2}m_{\chi }^{2}(11-24\gamma
_{M})}{2(m_{\phi }^{2}+2m_{\chi }^{2})}\right] .  \label{ptF}
\end{equation}%
which upgrades the first line of (\ref{ptpr}) and agrees with the result of 
\cite{ABP}.

The cosmic RG flow allows us to derive the spectral index to the
next-to-leading log order (in \cite{ABP} it was computed to the leading
order), as well as the whole running of the spectrum to the next-to-leading
log order. Using the beta function $\beta _{\alpha }=-2\alpha
^{2}-(5/3)\alpha ^{3}$ to order $\alpha ^{3}$, the right-hand sides of (\ref%
{runT}) are upgraded to%
\begin{eqnarray}
n_{T}=-\beta _{\alpha }(\alpha )\frac{\partial \ln P_{T}}{\partial \alpha }
&=&-\frac{12\alpha ^{2}m_{\chi }^{2}(1+4\alpha \gamma _{M})}{m_{\phi
}^{2}+2m_{\chi }^{2}}+\frac{12\alpha ^{3}m_{\chi }^{2}(m_{\phi
}^{2}+8m_{\chi }^{2})}{(m_{\phi }^{2}+2m_{\chi }^{2})^{2}},  \notag \\
\frac{m_{\phi }^{2}+2m_{\chi }^{2}}{2^{n}m_{\chi }^{2}\alpha ^{n+2}(n+1)!}%
\frac{\mathrm{d}^{n}n_{T}}{\mathrm{d}\ln k\hspace{0.01in}\hspace{0.01in}^{n}}
&=&\frac{m_{\phi }^{2}+2m_{\chi }^{2}}{2^{n}m_{\chi }^{2}\alpha ^{n+2}(n+1)!}%
\left( -\beta _{\alpha }(\alpha )\frac{\partial }{\partial \alpha }\right)
^{n}n_{T}  \label{runniTF} \\
&=&-12-24(n+2)\alpha \gamma _{M}+(n+2)\alpha \left( 3\frac{26m_{\chi
}^{2}+7m_{\phi }^{2}}{m_{\phi }^{2}+2m_{\chi }^{2}}-10\sum_{k=1}^{n+2}\frac{1%
}{k}\right) .  \notag
\end{eqnarray}

\subsection[Scalar fluctuations and first correction to r=-8nT]{Scalar
fluctuations and first correction to $r=-8n_{T}$}

\label{scalarocm}

It was shown in ref. \cite{ABP} that the fakeon $\chi _{\mu \nu }$ belonging
to the gravitational triplet does not affect the quantum gravity predictions
concerning the scalar fluctuations, up to the next-to-leading order
included. The results of this paper promote the same conclusion to the
next-to-leading log order, where they coincide with those of formulas (\ref%
{ptpr}) and (\ref{runS}). In particular,%
\begin{eqnarray}
\mathcal{P}_{\mathcal{R}}(k) &=&\frac{Gm_{\phi }^{2}}{12\pi \alpha (1/k)^{2}}%
\Big(%
1+(5-4\gamma _{M})\alpha (1/k)%
\Big)%
\qquad n_{\mathcal{R}}-1=-4\alpha +\frac{4}{3}\alpha ^{2}(5-6\gamma _{M}), 
\notag \\
\frac{\mathrm{d}^{n}(n_{\mathcal{R}}-1)}{\mathrm{d}\ln k\hspace{0.01in}%
\hspace{0.01in}^{n}} &=&-2^{n+2}\alpha ^{n+1}n!\left[ 1-\frac{n+1}{6}\alpha
\left( 15-12\gamma _{M}-5\sum_{k=1}^{n+2}\frac{1}{k}\right) \right] .
\label{runniSF}
\end{eqnarray}

Considering the \textquotedblleft dynamical\textquotedblright\
tensor-to-scalar-ratio%
\begin{equation}
r(k)=\frac{\mathcal{P}_{T}(k)}{\mathcal{P}_{\mathcal{R}}(k)},  \label{dynr}
\end{equation}%
our results allow us to compute the first correction to the relation $%
r=-8n_{T}$. Using (\ref{ptF}), (\ref{runniTF}) and (\ref{runniSF}), we find,
to the order $\alpha ^{3}$ included,%
\begin{equation}
r+8n_{T}=\frac{\mathcal{P}_{T}}{\mathcal{P}_{\mathcal{R}}}+8n_{T}=-\frac{%
384m_{\chi }^{2}\alpha (1/k)^{3}}{m_{\phi }^{2}+2m_{\chi }^{2}}.
\label{ratiodevia}
\end{equation}

We recall that the theory does not predict other degrees of freedom (when
the matter sector is switched off), because the fakeon projection eliminates
any additional scalar and tensor perturbations, as well as the vector
perturbations.

\section{Testable predictions}

\label{predictions}\setcounter{equation}{0}

In this section we work out a number of predictions that have a chance to be
tested in the incoming years \cite{CMBStage4}. We express the results in
terms of a \textquotedblleft pivot\textquotedblright\ scale $k_{\ast }$ and
evolve $\alpha (1/k)$ from $k_{\ast }$ to $k$ by means of the RG evolution
equations, using the next-to-leading log solution (\ref{runca}). So doing,
the spectra become functions of $\alpha _{\ast }\equiv \alpha (1/k_{\ast })$
and $\ln (k_{\ast }/k)$.

\begin{figure}[t]
\begin{center}
\includegraphics[width=7.5truecm]{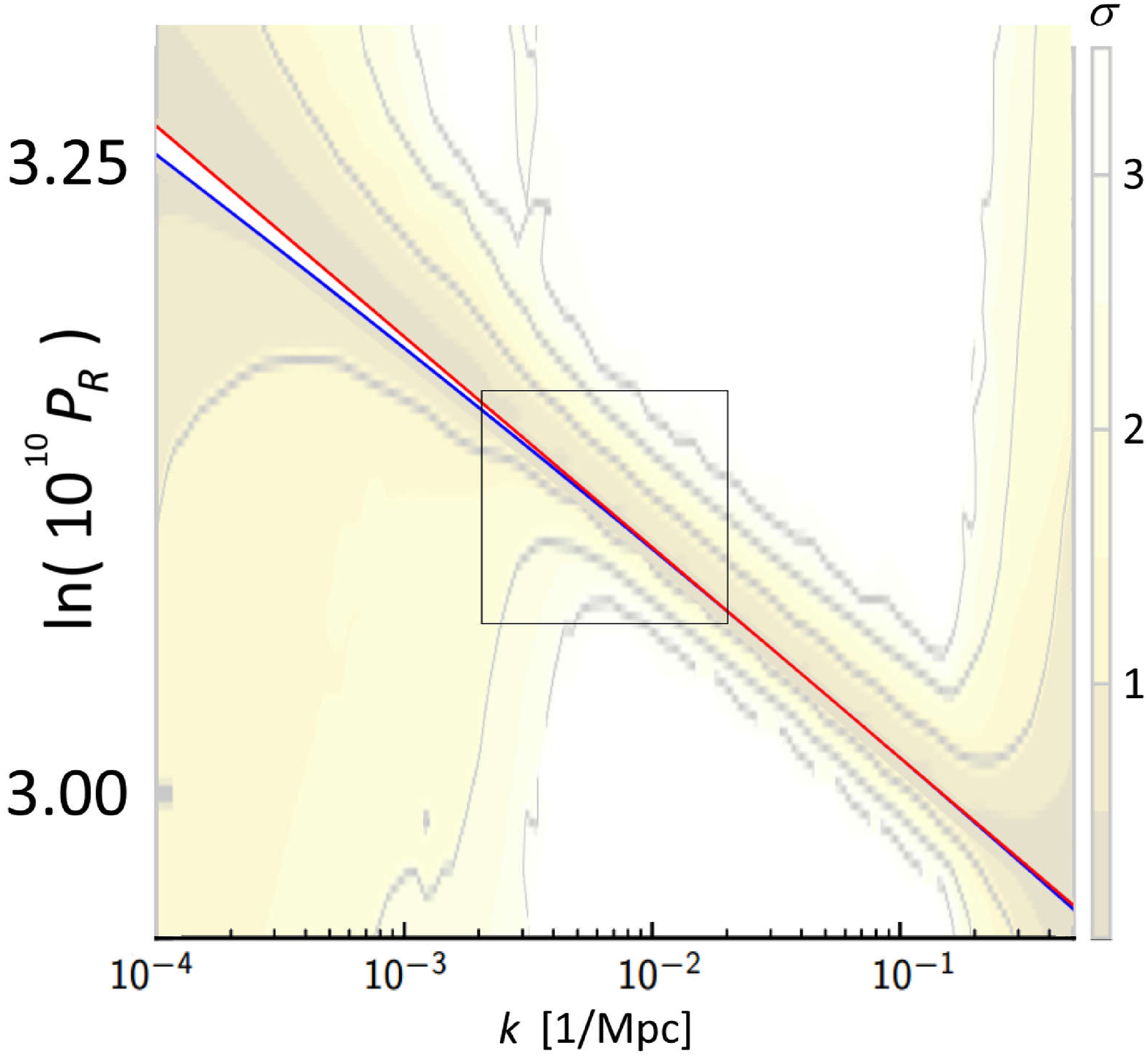}\quad %
\includegraphics[width=7.5truecm]{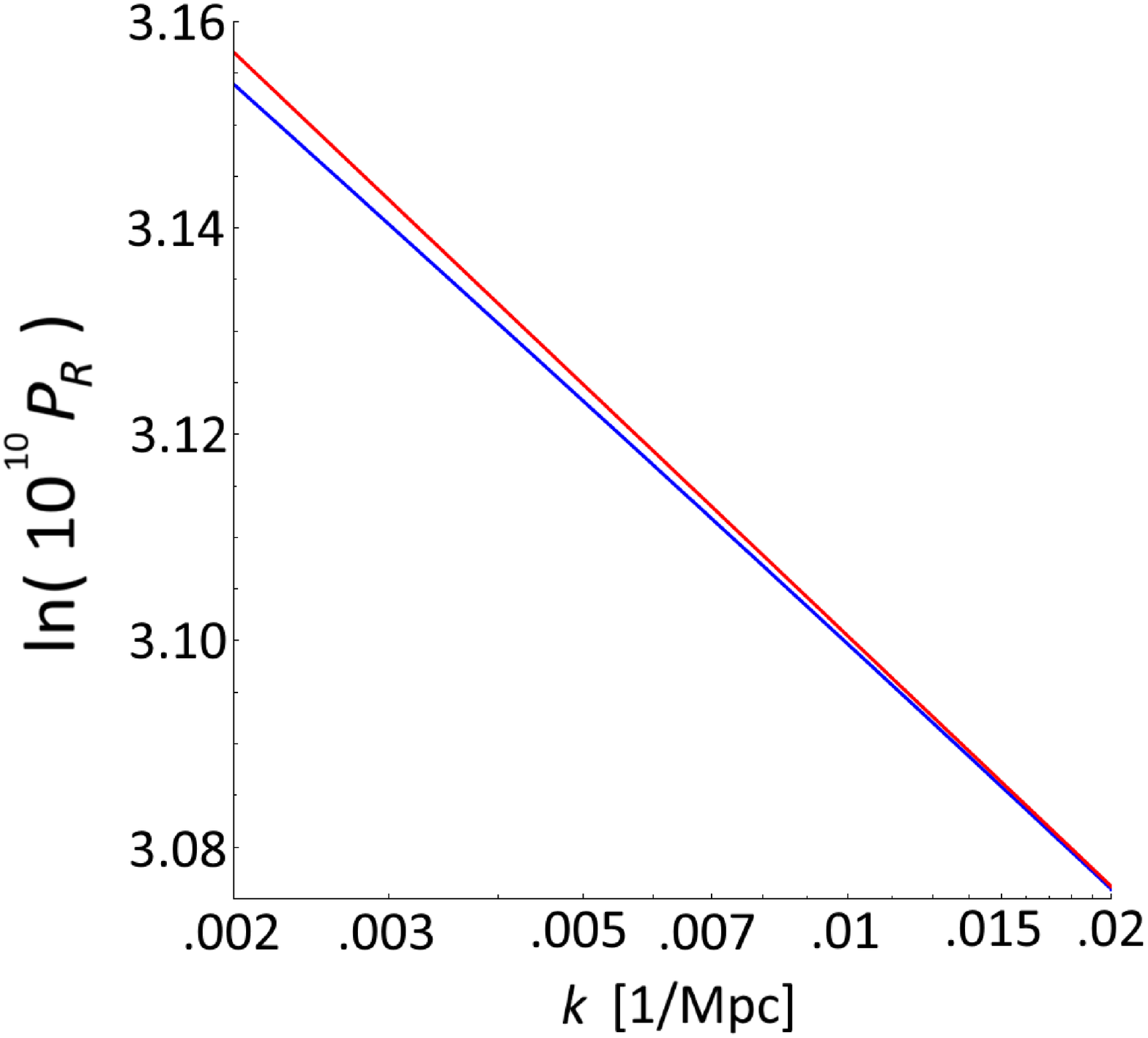}
\end{center}
\caption{Running spectrum of the scalar fluctuations (in blue) compared to
the nonrunning (linearized) truncation (in red). The left picture is
superposed to the data reported in \protect\cite{Planck18}. The right
picture magnifies the square at the center of the left picture}
\label{PlanckSpec}
\end{figure}

\begin{figure}[t]
\begin{center}
\includegraphics[width=14truecm]{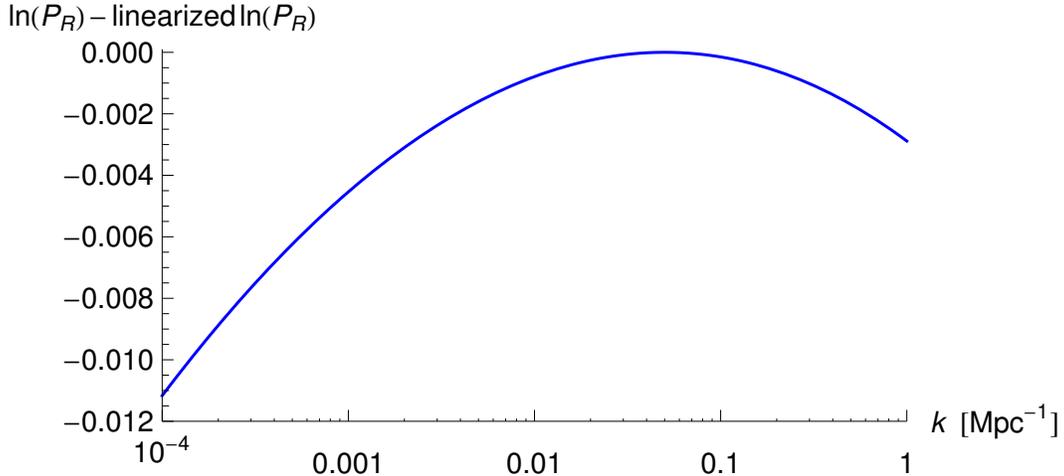}
\end{center}
\caption{Running of the scalar spectrum}
\label{RunningS}
\end{figure}

We take $k_{\ast }=0.05$ Mpc$^{-1}$ and plot the spectra in the range $%
10^{-4}$ Mpc$^{-1}\leqslant k\leqslant 1$ Mpc$^{-1}$. The data reported in 
\cite{Planck18} give%
\begin{equation*}
\ln (10^{10}\mathcal{P}_{\mathcal{R}}^{\ast })=3.044\pm 0.014,\qquad n_{%
\mathcal{R}}^{\ast }=0.9649\pm 0.0042,
\end{equation*}%
where the star superscript means that the quantity is evaluated at $k_{\ast
} $. Using formulas (\ref{runniSF}) we find%
\begin{equation*}
\alpha _{\ast }=0.0087\pm 0.0010,\qquad m_{\phi }=(2.99\pm 0.37)\cdot 10^{13}%
\text{GeV.}
\end{equation*}%
Taking the mean values just listed, the logarithm of the scalar power
spectrum $\mathcal{P}_{\mathcal{R}}(k)$ given in formula (\ref{runniSF})
plots as shown in the images of figure \ref{PlanckSpec}, which give an idea
of the precision we need to confirm the running experimentally. The blue
line is the running spectrum, while the red line denotes the linearized
truncation%
\begin{equation}
\left. \ln (10^{10}\mathcal{P}_{\mathcal{R}}(k))\right\vert _{\text{%
linearized}}\equiv \ln (10^{10}\mathcal{P}_{\mathcal{R}}^{\ast })+(n_{%
\mathcal{R}}^{\ast }-1)\ln \frac{k}{k_{\ast }}.  \label{linearPR}
\end{equation}%
The background of the left picture is an elaboration of the top-right image
of Fig. 20 of ref. \cite{Planck18}, which was obtained using Planck TT, TE,
EE + lowE + lensing. The right picture magnifies the square that appears at
the center of the left picture. The theoretical error is too small to be
plotted (the contributions we are neglecting are of order $\alpha _{\ast
}^{2}\simeq 0.01\%$). 
We emphasize the effects of the running in fig. \ref{RunningS}, by plotting
the difference 
\begin{equation*}
\ln \mathcal{P}_{\mathcal{R}}(k)-\left. \ln \mathcal{P}_{\mathcal{R}%
}(k)\right\vert _{\text{linearized}}.
\end{equation*}%
By construction, this difference vanishes at the pivot scale. 

\begin{figure}[t]
\begin{center}
\includegraphics[width=14truecm]{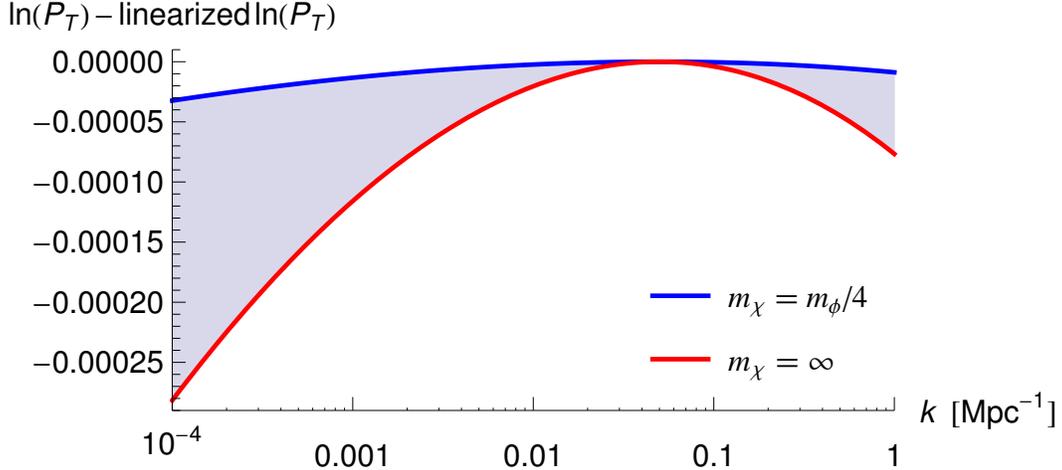}
\end{center}
\caption{Running of the tensor spectrum}
\label{RunningT}
\end{figure}

The running of the spectrum $\mathcal{P}_{T}(k)$ of the tensor fluctuations
is shown in fig. \ref{RunningT}, by plotting the difference%
\begin{equation*}
\ln \mathcal{P}_{T}(k)-\left. \ln \mathcal{P}_{T}(k)\right\vert _{\text{%
linearized}},\qquad \left. \ln \mathcal{P}_{T}(k)\right\vert _{\text{%
linearized}}\equiv \ln \mathcal{P}_{T}^{\ast }+n_{T}^{\ast }\ln \frac{k}{%
k_{\ast }}.
\end{equation*}%
The mass $m_{\chi }$ of the fakeon $\chi _{\mu \nu }$ is constrained to lie
in the range $m_{\phi }/4<m_{\chi }\,<\infty $ by the consistency of the
fakeon prescription/projection \cite{ABP}. The blue border of fig. \ref%
{RunningT} is the spectrum for $m_{\chi }\,=m_{\phi }/4$, while the red
border is the spectrum for $m_{\chi }\,=\infty $. 

\begin{figure}[t]
\begin{center}
\includegraphics[width=11truecm]{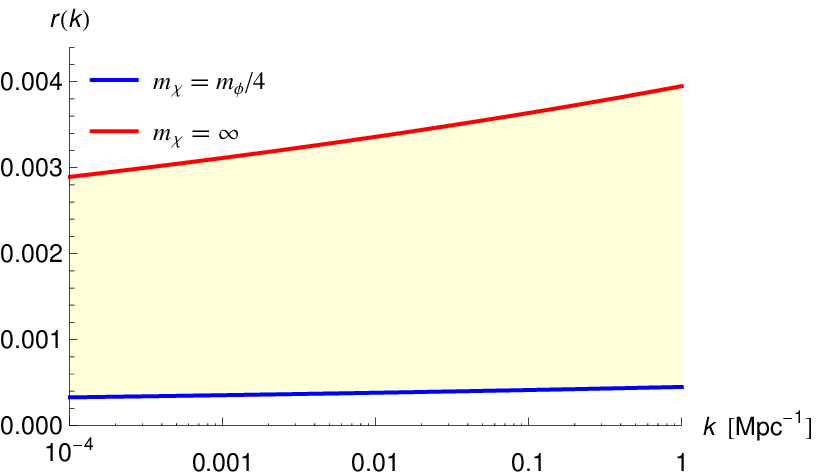}
\end{center}
\caption{Running of the tensor-to-scalar ratio $r(k)$}
\label{RunningR}
\end{figure}

Plotting the dynamical tensor-to-scalar ratio (\ref{dynr}) as a function of $%
k$, we obtain fig. \ref{RunningR}. 

At the pivot scale, the correction (\ref{ratiodevia}) to the relation $%
r=-8n_{T}$ is predicted to be%
\begin{equation*}
1000\left( r^{\ast }+8n_{T}^{\ast }\right) =\left\{ 
\begin{tabular}{l}
$-0.0146\pm 0.0049\qquad $ for $m_{\chi }=m_{\phi }/4,$ \\ 
$-0.131\pm 0.044$\qquad\ for $m_{\chi }=\infty .$%
\end{tabular}%
\right.
\end{equation*}

We recall that the measurement of one quantity, such as $r^{\ast }$ (or $%
n_{T}^{\ast }$), is enough to determine the only remaining free parameter of
quantum gravity with fakeons, which is $m_{\chi }$. After that, it is
possible to make precision tests of the other predictions.

\section{Conclusions}

\label{conclusions}\setcounter{equation}{0}

Quantum gravity with purely virtual particles allows us to anticipate the
outcomes of various measurements in inflationary cosmology and maybe pave
the way for precision tests. The constraints of quantum field theory
overcome the lack of predictivity due to the arbitrariness of classical
theories.

In this paper we laid out an important ingredient of the relation between
cosmology and high-energy physics by reformulating the history of the early
universe as the evolution of a flow that bears many resemblances with the RG
flow we are familiar with from quantum field theory. Instead of being driven
by the radiative corrections, it is due to the dependence on the background
FLRW\ metric. Although the two flows are very different in nature, many
properties of the slow roll expansion and the inflationary perturbation
spectra in the superhorizon limit are similar to the ones encoded into the
Callan-Symanzik equation satisfied by the amplitudes of quantum field
theory, at least from the mathematical and formal points of view.

The cosmic RG flow starts from a fixed point, which is de Sitter space,
around which the beta function behaves like the one of pure QCD. The flow
does not end into a fixed point, but continues into the reheating phase,
which we have not investigated here. Viewing inflation this way offers a
better understanding of it and allows us to simplify various computations,
upgrading the techniques that have been used earlier. It might also inspire
new, unforeseen developments. To some extent, it might be possible to
describe parts of the later evolution of the universe as RG flows with
modified beta functions, to take into account the contributions of matter.

The RG flow of inflation is of a \textquotedblleft pure\textquotedblright\
type, in the sense that it is solely governed by the beta function, since
the power spectra have no anomalous dimensions. The core information of a
spectrum is the constant $Q(0)$ of (\ref{Q0}), which is a power series in
the coupling $\alpha _{0}$ with no a priori relation to the beta function.
Because of this, $Q(0)$ is specific of the (scalar, tensor, etc.) spectrum
we consider. The RG information is the part controlled by the RG equation,
so it is universal (i.e., the same for every spectrum) and encoded in the
beta function. Recalling that in quantum field theory the core information
of a correlation function is an extremely complicated nonlocal expression
that can be calculated by means of Feynman diagrams, we can appreciate that
the cosmic RG flow has the simplest, yet nontrivial structure we can hope
for.

The running of the power spectra can be studied efficiently by organizing
the perturbative expansion in terms of leading and subleading logs and
resumming the powers of $\alpha _{0}\ln \eta $ altogether. We have worked
out the scalar and tensor spectra and their runnings in quantum gravity to
the next-to-leading log order, together with the first correction to the
relation $r+8n_{T}=0$. 

Note that the coupling $\alpha $ of inflation has a value ($\sim 1/115$ at
the pivot scale) that is close to the value of the QED fine structure
constant. This makes the theoretical errors negligible, once we include a
few orders of the expansion in powers of $\alpha $, as we have done here.
Depending on the experimental resolution that will be achieved in the
future, cosmology might turn into an arena for precision tests of quantum
gravity.

The procedures described in this paper can be pushed to higher orders. This
requires to implement the fakeon projection beyond what we have done so far,
by following the iteration procedure of ref. \cite{FLRW}. As shown there,
the result is expected to be an asymptotic series. Given that the fine
structure constant $\alpha _{\ast }$ of inflation is as small as $1/115$,
this does not jeopardize the predictions for quite a while. Moreover, at
some point high powers of $\alpha $ are outcompeted by the loop corrections
we have been neglecting so far, which requires to rethink the whole strategy
anyway.

\vskip8truept \noindent {\large \textbf{Acknowledgments}}

\vskip 1truept

We are grateful to E. Bianchi, D. Comelli, F. Fruzza, G. Signorelli and M.
Piva for helpful discussions.

\end{document}